\documentclass[useAMS,usenatbib]{mn2e}
\usepackage{graphicx}

\usepackage[fleqn]{amsmath}

\usepackage{color}
\usepackage{float}

\usepackage{siunitx}
\usepackage{caption}

\usepackage{amssymb}

\newlength{\widthsingle}
\newlength{\widthdouble}
\newlength{\widthbig}

\newcommand{\F}{\mathrm{F}}

\title[Radial gas motions in The HI Nearby Galaxy Survey]{Radial gas motions in The HI Nearby Galaxy Survey (THINGS)}

\author[T. M. Schmidt et al.]{Tobias. M. Schmidt$^{1,2}$\thanks{E-mail: tschmidt@mpia-hd.mpg.de},
Frank Bigiel$^{1}$, Ralf S.\ Klessen$^{1,3,4}$, W.J.G. de Blok$^{5,6,7}$
\\\\
$^{1}$Universit\"{a}t Heidelberg, Zentrum f\"ur Astronomie, Institut f\"ur Theoretische Astrophysik, Albert-Ueberle Str. 2, \\ ~~~~~69120 Heidelberg, Germany\\
$^{2}$Max-Planck-Institut f\"{u}r Astronomie, K\"{o}nigstuhl 17, 69117 Heidelberg, Germany\\
$^{3}$Kavli Institute for Particle Astrophysics and Cosmology, Stanford University, SLAC National Accelerator Laboratory, \\ ~~~~~Menlo Park, CA 94025, USA\\
$^{4}$Department of Astronomy and Astrophysics, University of California, 1156 High Street, Santa Cruz, CA 95064, USA \\
$^{5}$Netherlands Institute for Radio Astronomy (ASTRON), Postbus 2, 7990 AA Dwingeloo, the Netherlands
\\
$^{6}$Astrophysics, Cosmology and Gravity Centre, Department of Astronomy, University of Cape Town, Private Bag X3, \\ Rondebosch 7701, South Africa\\
$^{7}$Kapteyn Astronomical Institute, University of Groningen, PO Box 800, 9700 AV Groningen, the Netherlands
}

\begin{document}

\date{Accepted 2015 December 28.  Received 2015 December 7; in original form 2015 April 9}

\pagerange{\pageref{firstpage}--\pageref{lastpage}} \pubyear{2015}

\maketitle

\label{firstpage}

\begin{abstract}
The study of 21$\,$cm line observations of atomic hydrogen allows detailed insight into the kinematics of spiral galaxies. We use sensitive high-resolution VLA data from The HI Nearby Galaxy Survey (THINGS) to search for radial gas flows  primarily in the outer parts (up to $3\times r_{25}$) of  ten nearby spiral galaxies. Inflows are expected to replenish the gas reservoir and fuel star formation under the assumption that galaxies evolve approximately in steady state.
We carry out a detailed investigation of existing tilted ring fitting schemes and discover systematics that can hamper their ability to detect signatures of radial flows. 
We develop a new Fourier decomposition scheme that fits for rotational and radial velocities and simultaneously determines position angle and inclination as a function of radius. 
Using synthetic velocity fields we show that our novel fitting scheme is less prone to such systematic errors and that it is well suited to detect radial inflows in disks.
We apply our fitting scheme to ten THINGS galaxies and find clear indications of, at least partly previously unidentified, radial gas flows, in particular for NGC~2403 and NGC~3198 and to a lesser degree for NGC~7331, NGC~2903 and NGC~6946. The mass flow rates are of the same order but usually larger than the star formation rates. At least for these galaxies a scenario in which continuous mass accretion feeds star formation seems plausible. The other galaxies show a more complicated picture with either no clear inflow, outward motions or complex kinematic signatures.
\end{abstract}


\section{Introduction}

The question of how spiral galaxies manage to sustain star formation activity over cosmological times is key to our understanding of galaxy formation and evolution. Finding an answer is not easy, given that most observations point towards very short gas depletion timescales. For example, our Galaxy forms new stars at a rate of $2 - 4\,$M$_\odot\,$yr$^{-1}$ (\citealt{Naab2006}, \citealt{Adams2013}). Its gas mass is about $8 \times 10^9\,$M$_\odot\,$ (see \citealt{Ferriere2001} and \citealt{Kalberla2003}). If we assume a constant star formation rate \citep[SFR,][]{Binney2000}, then the remaining gas should be converted into stars within about $2 - 4$ Gyr. Similar values or order of only a few billion years are reported for many nearby disk galaxies (for observational evidence, see for example \citealt{Bigiel2008}, \citealt{Leroy2008}, or \citealt{Bigiel2011}; for theoretical arguments, see e.g.\ \citealt{Pflamm2009}). At higher redshifts of $z \approx 2$, the  depletion timescale can be even shorter.  Values of $0.6-1.5$\,Gyr have been inferred for the total gas \cite[e.g.][]{Genzel2010}, while the numbers for the molecular gas component can be as low as $0.5$\,Gyr \citep{Daddi2010}. In vigorously star forming merging systems, the observations suggest even shorter timescales, down to $0.1\,$Gyr. 

We note, however, that these are average numbers, with the data showing significant variations, both from galaxy to galaxy and within each galaxy
as a function of many parameters, not just column density (for further discussions, see  \cite[e.g.][]{Saintonge2011, Saintonge2012, Leroy2013, Shetty2013, Shetty2014a, Shetty2014b}).

If we discard the possibility that most spiral galaxies are observed at the verge of running out of gas, and instead assume that they evolve in quasi steady state, then this requires a supply of fresh gas delivered to the galaxy at a rate roughly equal to its star formation rate. Indeed, there is additional support for this picture. 

First, galactic accretion flows are a natural outcome of cosmological structure formation calculations. For example, \citet{Dekel:2009p1173} and \citet{Ceverino2010} argue that massive galaxies are continuously fed by steady, narrow, cold gas streams that penetrate through the accretion shock at the virial radius into the central galaxy \cite[see also][]{Marinacci2013}. Roughly three quarters of all galaxies forming stars at a given rate are fed by smooth streams \cite[see also][]{Agertz2009}, at least up to redshifts of $z\approx2$. The details of this process are still under debate, as they seem to depend on the numerical method employed and on the way gas cooling is implemented \cite[e.g.][]{Bird2013}. Nevertheless, even if most of the gas shocks at the virial radius and builds up a hot ionised halo, some fraction of it may cool down again and become available for disk accretion. As proposed by \citet{Peek2009} this gas could condense into higher-density clumps at the interface between halo and disk and then ``rain'' down onto the disk later in the process. This transport of matter converts potential energy into kinetic energy, and constitutes a source for the observed turbulence on smaller scales and thus contributes to controlling the star formation process in the galaxy \citep{MacLow2004}. Indeed, gas accretion has been proposed to be the main source of turbulence in the extended outer disk of galaxies, where the star formation activity is low and consequently stellar feedback cannot provide enough energy and momentum to explain the observed velocity dispersion in the H\,{\sc i} gas \citep{Klessen2010}. 

Second, indirect support for continuous accretion comes from the fact that the observed amount of atomic gas in the universe appears to be roughly constant, from a  redshift of $z \approx 3$ till today, while the stellar content continues to increase, suggesting that overall the H\,{\sc i} content of typical galaxies is continuously replenished from some external reservoir \citep{Hopkins2008,Prochaska2009}. 

Third, for our Galaxy at least, further evidence for an ongoing inflow of low-metallicity material  stems from the presence of deuterium in the solar neighbourhood \citep{Linsky2003} as well as in the Galactic Centre \citep{Lubowich2000}. As deuterium is destroyed in stars and as there is no other known source of deuterium in the Milky Way, it is thought to be pristine material of extragalactic origin \citep{Ostriker1975,Chiappini2002} falling on the Galaxy for the first time. 
 
If gas is accreted onto a disk galaxy from great distances, the new material is most likely delivered to large galactic radii. This is a consequence of angular momentum conservation, and it finds additional support from the fact that the outer disk simply has a larger cross section than the central regions. As most of the star formation occurs in the inner parts, one might expect an inward flow of gas through the disk. It is this radial mass flux that we aim to detect in this study.

In many galaxies neutral hydrogen (H\,{\sc i}) is found out to radii much larger than the optical disk \cite[e.g.][]{walter2008, Bigiel2010a, Bigiel2010b}. This gas provides us with the possibility to look for infall signatures. Indeed, the asymmetry commonly seen in the outer disks of galaxies are often considered to be the result of gas accretion \citep{bou05, jia99, ost89, Fraternali2008}, triggered either by recent minor mergers \citep{Zaritsky1997} or by tidal interaction \citep{kor02}. Some galaxies show one very prominent spiral arm extending beyond the rest of the H\,{\sc i} disk. Others have tails or a very asymmetric and lopsided disk or show substantial warps, especially in the outskirts. Also H\,{\sc i} bridges to companion galaxies are observed. \citet{Sancisi2008} argue that these observations are evidence for recent or ongoing accretion and infall of cold gas, and they claim that in at least 25\% of field galaxies indications for a merger event or tidal interactions can be detected. 

The detection of high velocity clouds (HVCs, \citealt{mil09}) points in a similar direction. These are extra-planar H\,{\sc i} clouds with low metallicities \citep{van04}, with anomalous kinematics \citep{Fraternali2002}, and with velocity gradients that are perpendicular to the Galactic plane \citep{fra05}. Many of them have a ''head-tail'' shape as is expected for the motion of colder and denser clouds through the tenuous hot halo gas of the Milky Way. Their origin, however, remains unclear. They could be fresh material falling onto the Galaxy for the first time. Or the HVCs could be the fallback of Galactic fountain flows \citep{Shapiro1976, Bregman1980}. In this case HVC's contain enriched gas that is ejected from the disk by stellar feedback. This gas is likely to fall back at larger radii. As it is lifted from the disk midplane, it runs down a vertical pressure gradient and some fraction of it spreads to larger radii. It travels through the hot halo gas above the plane that is rotating more slowly \citep{Fraternali2006, Fraternali2008}. As a consequence, the ejected material has larger angular momentum than its environment, and as it mixes with the extraplanar gas, it falls back to the disk at larger radii than it came from. 

Gas that accretes onto the outer disk must move inwards through this disk, before it becomes converted into stars in the star forming inner regions of the galaxy. To search for these radial gas flows, \citet{Wong2004} studied  seven spiral galaxies based on CO and H\,{\sc i} observations. They found elliptical and bar streaming motions and could give upper limits on the magnitude of possible radial inflows. However, without the availability of deep large-scale maps, the focus was on the inner disks, where the non-circular motions detected were of the order of $5 - 10 \;$km$\,$s$^{-1}$, comparable to the level of turbulence in the disk. Our approach is different. We focus on the extended outer disks of  nearby spiral galaxies, where the gas density is low and therefore the local streaming velocity needs to be high in order to result in an appreciable mass flow. 

We base our analysis on 21~cm line observations of neutral hydrogen in The H\,{\sc i} Nearby Galaxy Survey (THINGS,  \citealt{walter2008}). This allows to trace gas up to very large radii, typically far beyond the optical radius of the galaxy. The survey provides two-dimensional high-resolution velocity maps, and our approach to model these velocity fields is in general known as ''tilted ring fit''. The map is decomposed into individual rings which are circular within the plane of the disk, but inclined with respect to the line-of-sight and therefore appear elliptical. For each of them, inclination, position angle, circular velocity, and other parameters are determined.
The method was first described by \citet{Rogstad1974} and developed further by e.g. \citet{Bosma1978} and \citet{Begeman1987}. These studies derived rotation curves for a number of galaxies based on H\,{\sc i} maps from the Westerbork Synthesis Radio Telescope. Other studies extended this concept by also fitting non-circular velocity components of different kinds to the velocity field.

High-order harmonic decompositions were used by 
\citealt*{Schoenmakers1997} (applied to HI observations of NGC~2403 and NGC~3198) 
and 
\citealt*{Trachternach2008} (search for elongated potential in 19 THINGS galaxies).
Kinematic studies assuming flat disks and models of particular streaming motions were conducted by 
\citealt*{Wong2004} (search for radial inflow based on CO and HI observations of seven spiral galaxies),
\citealt*{Spekkens2007} (bar-like streaming motions in NGC~2976 based on H$\alpha$ and CO observations; development of the \texttt{velfit}-code\footnote{http://www.physics.rutgers.edu/$\sim$spekkens/velfit/})
and 
\citealt*{Sellwood2010} (improvement and extension of the previous study to five other galaxies based on H$\alpha$ and HI measurements).
Spekkens and collaborators later developed \texttt{DiskFit} \citep{Sellwood2015}\footnote{http://www.physics.rutgers.edu/$\sim$spekkens/diskfit/}, an improved version of \texttt{velfit} which is not strictly limited to constant inclination and can also operate on photometric data. \citet{Kuzio2012} applied \texttt{DiskFit} to H$\alpha$, CO and H\,{\sc i} kinematic data of NGC~6503.

Models with constant inclination are not suitable for studies including the regime beyond the star forming disk, where the H\,{\sc i} disks often show variable inclination. Furthermore, high-order Fourier decompositions would require the disk geometry to be determined in a separate, prior step by a low-order tilted ring fit.
In this study, we develop a fitting scheme that circumvents the complications associated with this two-step approach and at the same time can be applied to disks with (moderately) varying inclinations.

Our study is organised as follows. In Section~\ref{Data} we describe the THINGS data set. In Section~\ref{Theory} we outline our new method, discuss its stability properties, and compare to previous approaches. In Section~\ref{Verification} we apply our method to mock data, investigate its stability properties, and discuss its limitations. We apply the method to a sample of 10 galaxies from the THINGS survey and present our results in Section~\ref{Results}. Finally, we summarise and conclude in Section~\ref{sec:conclusions}.

\section{Data}
\label{Data}

\begin{table*}
\begin{center}

\parbox{0.8\textwidth}{
\caption{Properties of galaxies analysed}
\label{table1}

\begin{tabular}{cc S[table-format=3.4]S[table-format=3.4] S[table-format=2.1] S[table-format=1.2]S[table-format=2.2] S[table-format=3.1]S[table-format=1.2]S[table-format=2.2]@{$\;\pm$}S[table-format=1.1]} \hline
Name 	& Type 	& \multicolumn{1}{c}{Ra}  & \multicolumn{1}{c}{Dec}	& \multicolumn{1}{c}{Distance}	& r$_{25}$	& r$_{25}$	& M$_\mathrm{HI}$	& SFR	& \multicolumn{2}{c}{$\mathrm{\Gamma_{HI}(r>r_{25})}$} \\
 	& 	& $\mathrm{{}^{\circ}}$	& $\mathrm{{}^{\circ}}$	& $\mathrm{Mpc}$	& $\mathrm{{}^{\prime}}$	& $\mathrm{kpc}$	& $\mathrm{10^{8}\:M_{\odot}}$	& $\mathrm{M_{\odot}\,yr^{-1}}$	& \multicolumn{2}{c}{$\mathrm{M_{\odot}\,yr^{-1}}$} \\ \hline
NGC 2403	& SBcd	& 114.2129	& 65.6008	&  3.2	& 7.92	&  7.38	& 25.8	& 0.38	& -1.20 & 0.1  \\
NGC 2841	& Sb	& 140.5108	& 50.9764	& 14.1	& 3.46	& 14.19	& 85.8	& 0.74	&  0.36 & 0.5 \\
NGC 2903	& SBbc	& 143.0421	& 21.5011	&  8.9	& 5.87	& 15.21	& 43.5	& 0.44	& -0.54 & 0.1 \\
NGC 3198	& SBc	& 154.9792	& 45.5497	& 13.8	& 3.23	& 12.96	& 101.7	& 0.93	& -1.02 & 0.1 \\
NGC 3521	& SBbc	& 166.4525	& -0.0358	& 10.7	& 4.16	& 12.94	& 80.2	& 2.10	& -0.91 & 0.8 \\
NGC 3621	& Sd	& 169.5688	& -32.8142	&  6.6	& 4.89	&  9.38	& 70.7	& 2.09	&  0.59 & 0.9 \\
NGC 5055	& Sbc	& 198.9550	& 42.0292	& 10.1	& 5.87	& 17.26	& 91.0	& 2.12	&  1.35 & 0.5 \\
NGC 6946	& SBcd	& 308.7175	& 60.1539	&  5.9	& 5.74	&  9.85	& 41.5	& 3.24	& -13.12 & 2.0 \\
NGC 7331	& SAb	& 339.2671	& 34.4158	& 14.7	& 4.56	& 19.50	& 91.3	& 2.99	& -1.03 & 0.7 \\
NGC 0925	& SBd	& 36.8187	& 33.5789	&  9.2	& 5.36	& 14.34	& 45.8	& 0.56	&  2.47 & 0.5 \\ \hline
\end{tabular}\\[\smallskipamount]
Coordinates, types, distances, $\mathrm{r_{25}}$ (the radius where the galaxy B-band surface brightness drops below 25\;mag\:arcsec$^{-2}$) and H\,{\sc i} masses as measured by THINGS are from \citealt{walter2008}, SFR are from \citealt{Leroy2008}, except for NGC~2903 \citep{Popping2010} and NGC~3621 \citep{walter2008}.
The last column gives our measurement of the average radial mass flow rate outside of $\mathrm{r_{25}}$. Negative values denote inflow, positive outflow. The given uncertainties are the average errors of the radial H\,{\sc i} mass flow outside of $r_{25}$. 
}
\end{center}
\end{table*} 

The main observational basis of this study are high spatial resolution observations of atomic hydrogen for a large set of nearby galaxies. These are taken from The HI Nearby Galaxy Survey (THINGS, \citealt{walter2008} \footnote{The HI Nearby Galaxy Survey:\\http://www.mpia-hd.mpg.de/THINGS}). It is a large survey of 34 nearby galaxies in the 21~cm line of neutral hydrogen and was conducted at the NRAO Very Large Array (VLA). The survey provides a high spatial resolution of 6" to 12" and a channel width of $5\;\mathrm{km\:s}^{-1}$.

Not all galaxies targeted within the THINGS survey are suitable for our analysis. Objects have to be well resolved (large angular size), have intermediate inclinations (above $30^{\circ}$ to $40^{\circ}$) and show a relatively regular velocity field. We applied our method to all promising candidates and in the end, our analysis yielded useful results for 10 galaxies which are listed in Table~\ref{table1}.

Our targets have distances between 3 and 15~Mpc which results in a physical resolutions between 100 and 500~pc. The survey provides a detailed census of the H\,{\sc i} distribution in a large number of nearby galaxies at high angular resolution as is thus well suited for kinematic studies.

For this work we use various data products provided by the survey.  The intensity (zeroth moment) maps reflects the total H\,{\sc i} column density and the first moment map the H\,{\sc i} velocity field. Data cubes and moment maps are available in two version. They differ in the weighting scheme applied to the interferomeric data from individual antenna pairs during image reconstruction. The \textsc{robust} scheme is optimised for maximising resolution and contrast by giving long baselines a higher impact. We choose the data products assuming \textsc{natural} weighting. This scheme is more suitable for us since since it provides better signal-to-noise and better recovers diffuse emission. A detailed description of the survey and the data processing can be found in \citet{walter2008}.

Additionally, we also use velocity maps derived by fitting Hermite functions directly to spectra in THINGS data cubes. They should better reflect the bulk motion of the gas but require data with slightly higher signal-to-noise ratio compared to first moment maps.

\label{EBHIS}
There is always the possibility that interferometric observations miss flux at large spatial scales due to incomplete $uv$-coverage. To minimise this issue, THINGS includes observations in the compact VLA-D configuration. For verification, we compare the THINGS intensity maps to single dish observations from the Effelsberg-Bonn HI Survey (EBHIS) \citep{Kerp2011} and find no significant discrepancy in total fluxes for the regions of the disks we model in this study. A more detailed analysis by Daniel Lenz (AIfA Bonn, private communication) came to the similar conclusion that usually no more than 10\% flux is missing in the THINGS maps, at least as long as the galaxy fits within the VLA primary beam.

Optical observations are used to find the nearside of our target galaxies which, in combination with the velocity information, allows us to uniquely determine the direction of rotation and to break the degeneracy between inflow and outflow. The optical images are taken from the NASA/IPAC Extragalactic Database\footnote{http://ned.ipac.caltech.edu}.
To compare our inflow results to the average recent star formation rate, we use GALEX far UV (FUV) images. We use the final data products available from the GALEX archive\footnote{http://galex.stsci.edu/}.

\section{Theory and method}
\label{Theory}

In the following Sections~\ref{TiltedRingAnalysis} and \ref{FourierDecomposition} we introduce basic theoretical concepts regarding modelling 2D velocity fields of disk galaxies and describe the different fitting procedures we evaluated. Sections~\ref{TwoStepFit} and \ref{SimultaneousFit} compare sequential versus simultaneous fitting approaches and their impact on derived quantities. The optimum procedure we finally adopted is described in Section~\ref{Fit_Procedure}

\subsection{Tilted ring analysis}
\label{TiltedRingAnalysis}

The quantity accessible to observation is the line-of-sight velocity ($V_\rmn{los}$). To infer the rotational and radial velocities of the gas in the disk from spatially resolved measurements of $V_\rmn{los}$, one has to utilise a specific model that can be fitted to the data.
The straightforward approach is to assume a velocity field which is symmetric with respect to the disk centre and projected under some viewing geometry. The measured $V_{\rmn{los}}$ then has the form:
\begin{equation}
\label{Eq1}
V_{\rmn{los}} = V_{\rmn{sys}} + V_{\rmn{rot}} \, \sin(i) \, \cos(\theta) + V_{\rmn{rad}} \, \sin(i) \, \sin(\theta)
\end{equation}

Here $V_{\rmn{sys}}$ denotes the systemic velocity of the galaxy with respect to the observer. $V_{\rmn{rot}}$ is the rotation velocity and $V_{\rmn{rad}}$ is a velocity component in radial direction in the plane of the disk. The other variables are the inclination $i$ of the disk and the azimuthal angle $\theta$ which together with the radius $R$ forms a polar coordinate frame within the plane of the galaxy. The disk is assumed to be thin and no motions perpendicular to the plane of the galaxy are taken into consideration. 

All quantities mentioned above can in general be treated as functions of the radius $R$. In the analysis, the disk is basically sliced in circular annuli and the above decomposition of $V_{\rmn{los}}$ done for each ring individually which leads to the term ''tilted ring'' analysis. 

The line-of-sight velocity is a function of the sky coordinates while the terms on the right hand side of Equation~\ref{Eq1} are functions of the coordinates within the plane of the disk. Therefore, one needs a coordinate transformation between sky coordinates ($X$,~$Y$) and the disk coordinates ($R$,~$\theta$). To do this transformation, the geometry of the disk has to be known. It can be described by the centre coordinates ($X_\rmn{C}$,~$Y_\rmn{C}$), the inclination ($i$) and the position angle of the apparent major axis of the disk ($PA$) as measured in the plane of the sky. One representation of the transformation equations is the following:

\begin{equation}
 \label{trafo1}
 \cos(\theta) = \frac{-(X-X_\rmn{C})\sin(PA) + (Y-Y_\rmn{C})\cos(PA)}{R}\;,
\end{equation}
\begin{equation}
 \label{trafo2}
 \sin(\theta) = \frac{-(X-X_\rmn{C})\cos(PA) - (Y-Y_\rmn{C})\sin(PA)}{R \cos(i)}\;.
\end{equation}

This transformation is not straightforward to solve for $R$ and $\theta$ since the parameters $X_\rmn{C}$, $Y_\rmn{C}$, $PA$ and $i$ implicitly depend on $R$, if one describes the disk as sequence of independent rings with possibly different inclinations, position angles, and centre coordinates. In this case, the global transformation from sky coordinates to disk coordinates is in general neither unique nor continuous. This is, because the rings can overlap or gaps between them can appear. The only reason this transformation can be used in practice is, because $i$ and $PA$ usually vary smooth and slowly. Or to be more specific, the analysis has to be restricted to galaxies where this condition is fulfilled.

\begin{figure}
 \begin{center}
	\includegraphics[width=.80\widthsingle]{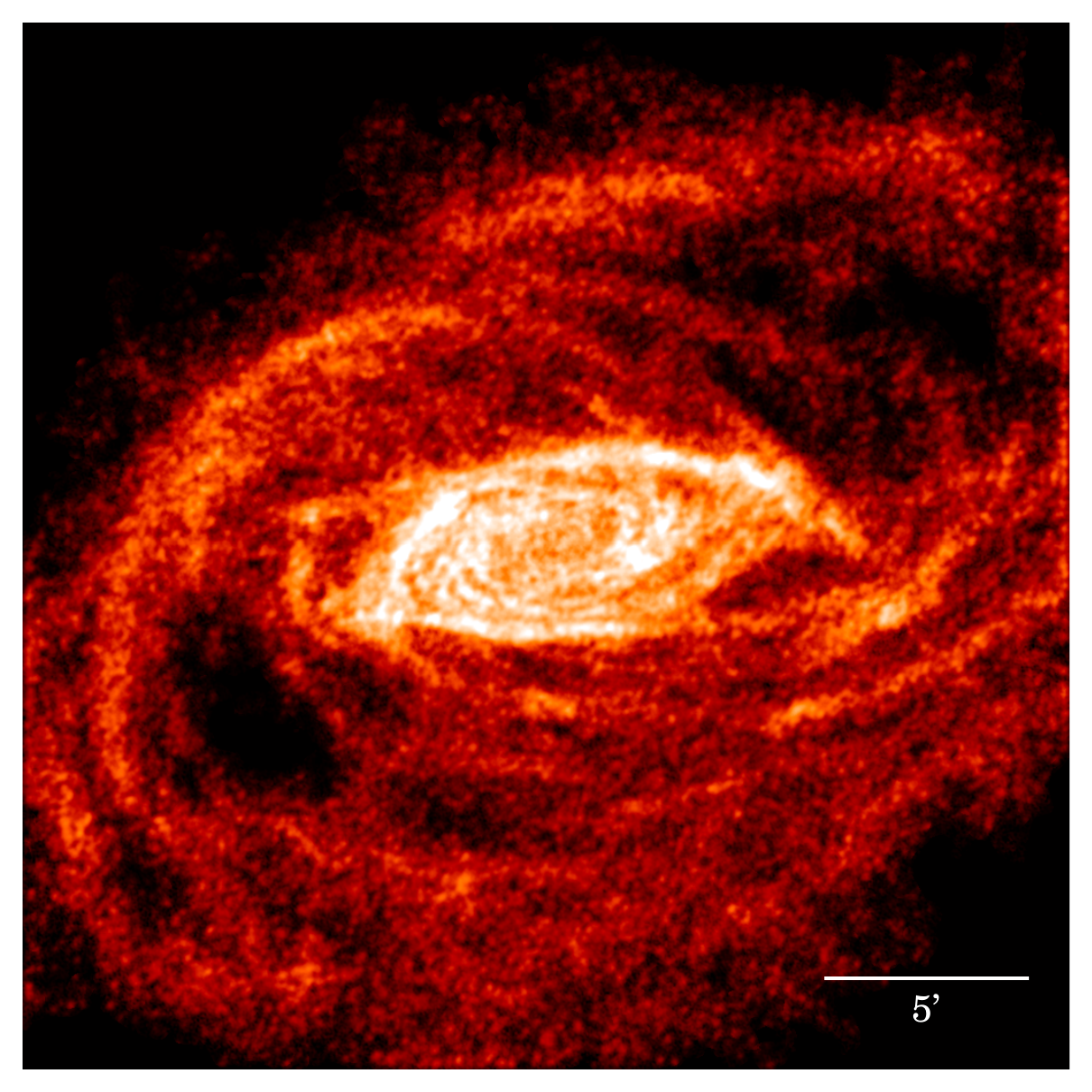}\\
	\includegraphics[width=.80\widthsingle]{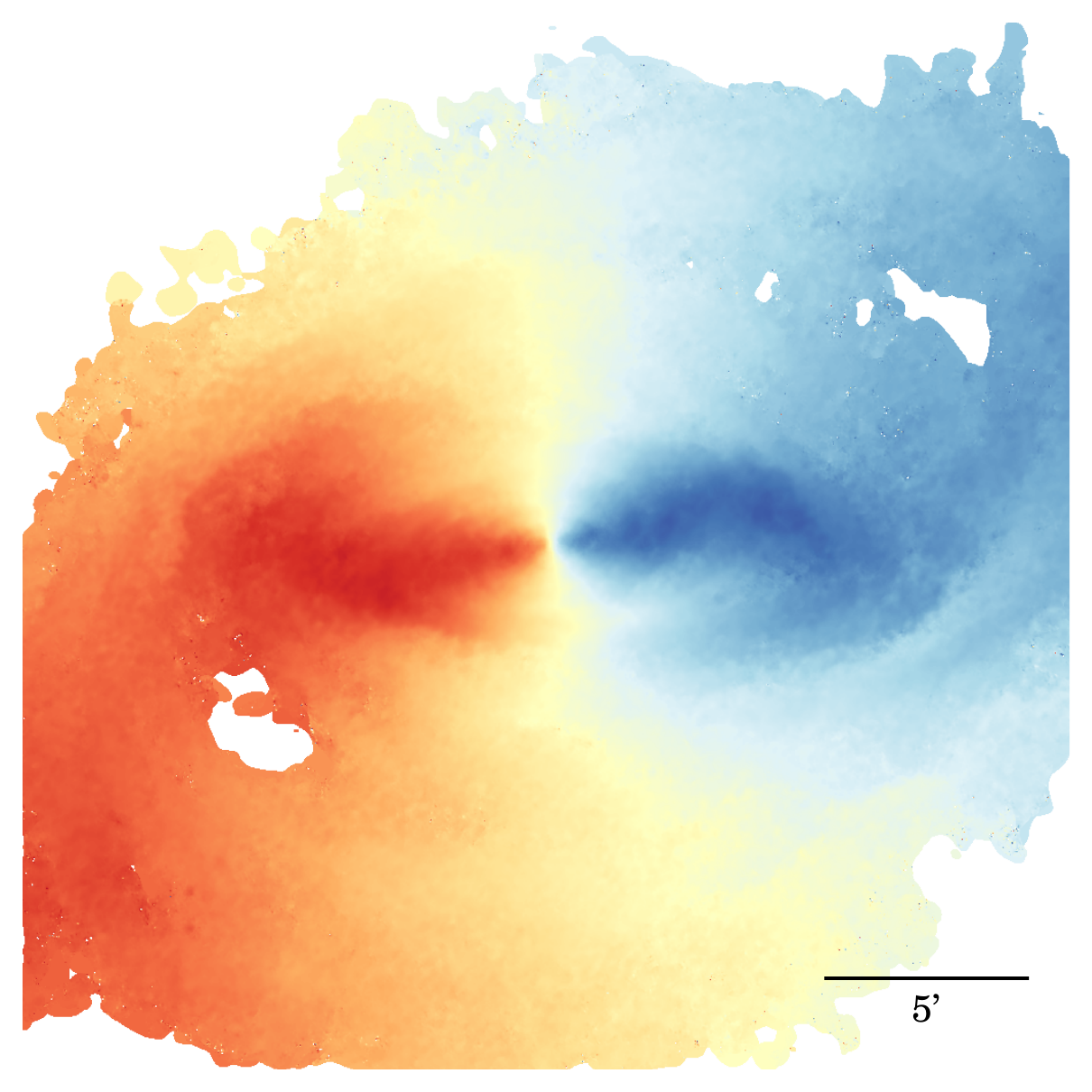}\\
	\includegraphics[width=.80\widthsingle]{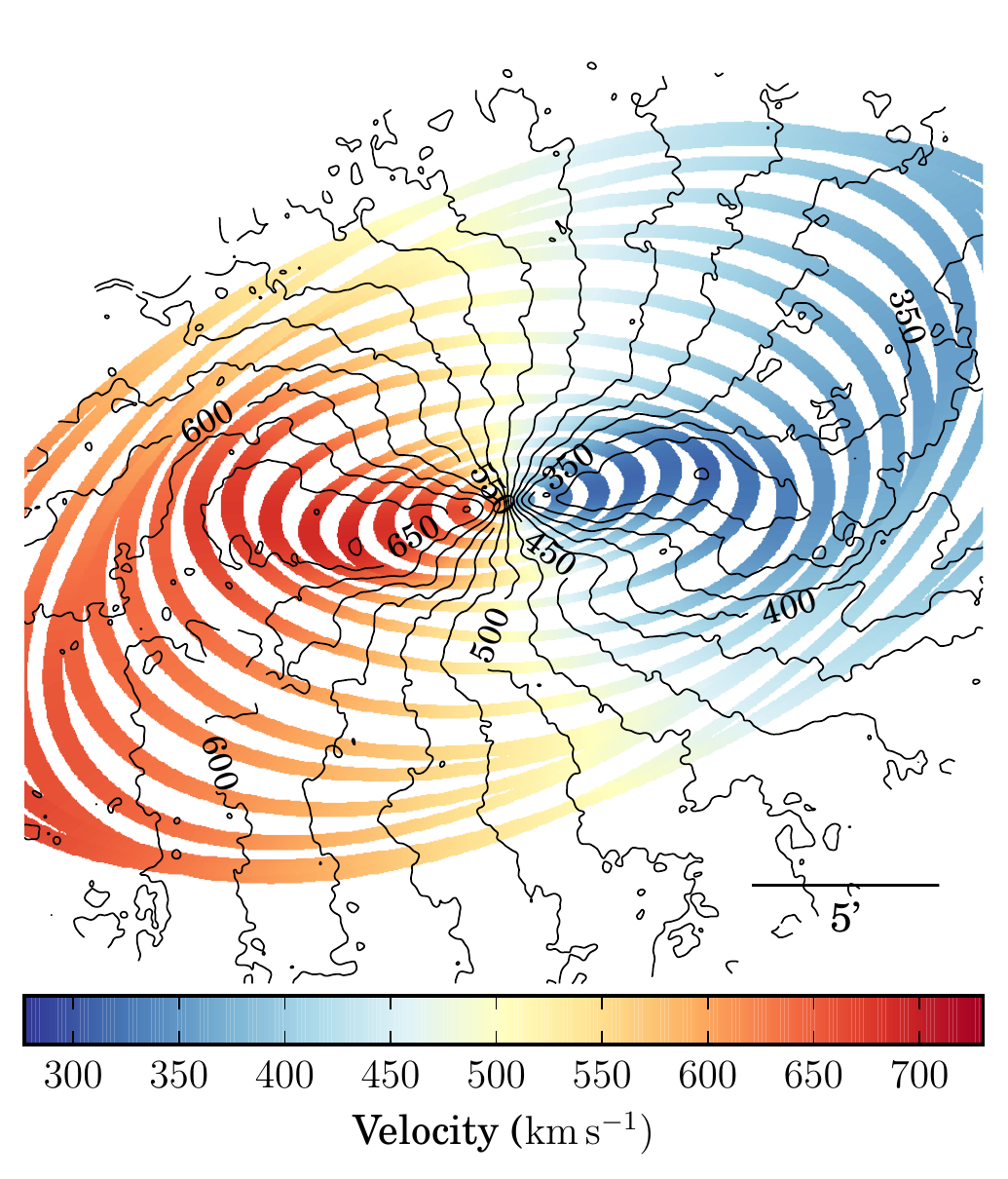}
	\caption{
	\textit{Top and centre:} Zeroth and first moment maps of NGC~5055. For the other galaxies, similar images can be found on the THINGS website: http://www.mpia-hd.mpg.de/THINGS\newline
	\textit{Bottom:} Illustration of a tilted ring fit to the velocity field of NGC~5055. For better visualisation only every second ring is shown. The colour scale represents the line-of-sight velocity
	and overplotted are the THINGS first moment map velocity contours in steps of $25\;\mathrm{km\:s}^{-1}$.
	}
	\label{fit_illustration_NGC5055}
 \end{center}
\end{figure}

An illustration of the analysis is given in Figure~\ref{fit_illustration_NGC5055}. It shows the observed H\,{\sc i} velocity of NGC~5055 as a "spider" diagram, where the lines denote regions of constant line-of-sight velocity, and in colours it indicates the corresponding tilted ring fits. It is clearly noticeable that NGC~5055 has a warped disk, but we also see how the tilted ring fit adapts to the varying disk geometry and follows the changing inclination and position angle.

Usually, not all parameters are kept free in a fit. In some schemes like the one implemented in \texttt{velfit} \citep{Spekkens2007} it is assumed that the disk is flat, and therefore all radii have the same inclination and position angle. In such cases, $i$ and $PA$ are determined once for the whole galaxy and then kept fixed for each radius.  Because we include the outer parts of the disks in our analysis which are often warped to some degree, we allow $PA$ and $i$ to vary independently for each ring. Only considering average values of $i$ and $PA$ would not  represent the actual geometry of the disk to the necessary precision.

However, leaving all fit parameters ($PA$, $i$, $X_\rmn{C}$, $Y_\rmn{C}$, $V_{\rmn{rot}}$, ...) completely unconstrained often leads to unstable fit results. We therefore choose to keep the ring centres fixed, assuming that all parts of the galaxy rotate around a common centre-of-mass. This is a good approximation for galaxies that are not subject to strong external perturbations, such as exerted by a massive satellite galaxy. We determine the galaxy centre from the inner, mostly flat part of the disk and enforce this centre coordinates for all rings.

\subsection{Fourier decomposition}
\label{FourierDecomposition}

By including the $V_{\rmn{rad}}$ term, the tilted ring fit as described above, already includes non-circular motions. Also all parameters especially inclination and position angle can be allowed to vary from ring to ring and therefore with radius. However, the velocity field of each ring is still restricted to rotational symmetry around the ring centre. One can generalise the model by allowing $V_{\rmn{rot}}$ and $V_{\rmn{rad}}$ to vary with respect to the azimuthal angle $\theta$. Again, it is not possible to allow arbitrary functions here, since not enough observables are available to determine them. The usual approach is to expand the azimuthal dependence of $V_{\rmn{rot}}$ and $V_{\rmn{rad}}$ in a Fourier series and include terms up to a certain order:
\begin{equation}
	\label{Eq2}
	V_{\rmn{rad}}(R,\theta)	= B_0 +	\sum_{k>0} A_k \sin(k \theta) + B_k \cos(k \theta)\;,
\end{equation}
\begin{equation}
	\label{Eq3}
	V_{\rmn{rot}}(R,\theta)	= D_0 +	\sum_{k>0} C_k \sin(k \theta) + D_k \cos(k \theta)\;.
\end{equation}
The line-of-sight velocity has to be expanded in the same way, and for notational convenience a factor $\sin(i)$ is already pulled out of all Fourier parameters:
\begin{equation}
	\label{Eq4}
	V_{\rmn{los}}(R,\theta)	= V_{\rmn{sys}} + \sin(i) \Big[ c_0 +	\sum_{k>0} s_k \sin(k \theta) + c_k \cos(k \theta)\Big]\;.
\end{equation}
We denote Fourier parameters corresponding to the velocities in the disk frame by capital letters, while the ones referring to the line-of-sight velocity are lower case. 

By inserting equations~\ref{Eq2} and \ref{Eq3} in Equation~\ref{Eq1} and comparing with Equation~\ref{Eq4}, one can find relations between the Fourier coefficients of the line-of-sight velocity and the Fourier parameters of the velocities in the disk frame:
\begin{equation}
 \label{Eq5}
 \begin{split}
	V_{\rmn{los}}	&	=	V_{\rmn{sys}} + \frac{1}{2} \sin(i) \, \Big[	\, (D_1+A_1)			\\
								& + ( 2 B_0 + C_2 - B_2) \sin(\theta)															\\
								& +	( 2 D_0 + D_2 + A_2) \cos(\theta) 														\\
								& + \sum_{k>1} (C_{k-1}+C_{k+1}+B_{k-1}+B_{k+1}) \sin(k \theta)		\\
								& + \sum_{k>1} (D_{k-1}+D_{k+1}-A_{k-1}+A_{k+1}) \cos(k \theta) 	\, \Big]\;.
 \end{split}
\end{equation}
One finds that a radial or rotational velocity component of order $k$ contributes to order $k-1$ and $k+1$ to the line-of-sight velocity. Also, there are roughly two times as many Fourier coefficients in the disk frame ($A_k$, $B_k$, $C_k$, $D_k$) than are accessible by observations ($s_k$, $c_k$). This makes a unique reconstruction of the actual velocities impossible. 

We therefore restrict our analysis to second order in $V_{\rmn{los}}$, corresponding to first order in $V_{\rmn{rot}}$ and $V_{\rmn{rad}}$. This simplifies Equation~\ref{Eq5} to:
\begin{equation}
 \label{Eq8}
	\begin{split}
	V_{\rmn{los}}	& =	V_{\rmn{sys}} + \frac{1}{2} \sin(i) \, \Big[ 	( D_1 + A_1 )																																\\
								&	+																							( 2 B_0 ) \sin(\theta)						+	( 2 D_0) \cos(\theta) 									\\
								&	+																							( C_{1} + B_{1} ) \sin(2 \theta)	+ ( D_{1} - A_{1} ) \cos(2 \theta) 	\, \Big]\;.
  \end{split}
\end{equation}
In this simplified model it is easier to solve for the Fourier coefficients of $V_{\rmn{rot}}$ and $V_{\rmn{rad}}$. The comparison to Equation~\ref{Eq4} shows that the $V_{\rmn{los}}$ parameter $c_1$  directly represents the rotation curve and the $s_1$ parameter corresponds to the 0th order radial velocity, sometimes also denoted as $V_{\rmn{rad,0}}$. There are only two remaining degeneracies. In principle the 0th order term $c_0 = \frac{1}{2} (D_1+A_1)$ is degenerate with the systemic velocity. However, this is not a significant concern, because the systemic velocity should be the same for all rings and therefore needs to be determined only once. This can, for example, be done by directly using the velocity of the galaxy centre. A degeneracy that cannot be solved concerns  $s_2 = \frac{1}{2}(C_{1}+B_{1})$, because $C_1$ and $B_1$ only appear in this term. To make progress, we adopt $C_1 = B_1$. That  is, we assume that both terms contribute equally. 

Within a single ring, all parameters are modelled without any radial dependence and therefore with constant rotation velocity. This prohibits a perfect fit to the data in the inner part of a galaxy with roughly linearly rising rotation curve. Because the focus of this study are not the galactic centres, and because spiral galaxies show a mostly flat rotation curve over most of their outer disk, this is not a concern for this study.

\label{solid_body}
Solid-body rotation produces another problem. 
It leads to parallel iso-velocity contours in the observed velocity field and makes it impossible to disentangle $V_\rmn{rot}$ and $\sin(i)$. Therefore our analysis cannot be applied to dwarf galaxies which usually have linearly rising rotation curves over most of their H\,{\sc i} disk. The same effect causes problems in the inner parts of spiral galaxies. Since these parts are not the focus of our study we disregard this regime when applying our analysis and interpreting our results.

\subsection{Two-step fit}
\label{TwoStepFit}
The Fourier components themselves form an orthogonal base and therefore can be fitted independent of each other. However, at least the components of odd order couple to inclination and position angle. It is therefore not obvious that one can first determine the ring geometry with a tilted ring fit of low order, and then in a second step, apply a higher order Fourier decomposition using the previously determined ring geometry. 

Because this is an often-used procedure, implemented e.g. by the \texttt{reswri} tasks (which uses \texttt{ROTCUR} as the first step) within the \texttt{GIPSY}\footnote{Groningen Image Processing System: \\ http://www.atnf.csiro.au/computing/software/gipsy/index.html} package, we decided to test the applicability of this approach. We follow a two-step procedure: First, a tilted ring analysis of first order is performed, where we allow inclination and position angle to vary fully independently for each ring.  In a second step, a Fourier decomposition of the velocity field is calculated, in which inclination and position angle are fixed to the values determined in the tilted ring fit. We compare two slightly different methods. In the first one, the initial tilted ring fit already includes radial motions in the form of the $V_\rmn{rad}$ term in Equation~\ref{Eq2}. In the second one, it is restricted to a purely rotational velocity field. The latter is similar to the one used by \cite{Schoenmakers1997} and \cite{Trachternach2008}. Technically, they performed the Fourier decomposition not on the full velocity field, but instead on the residuals of the tilted ring fit. Nonetheless, in this approach it means the ring geometry is derived from a purely rotational tilted ring fit.

\begin{figure*}
 \begin{center}	
	\includegraphics[width=\widthdouble]{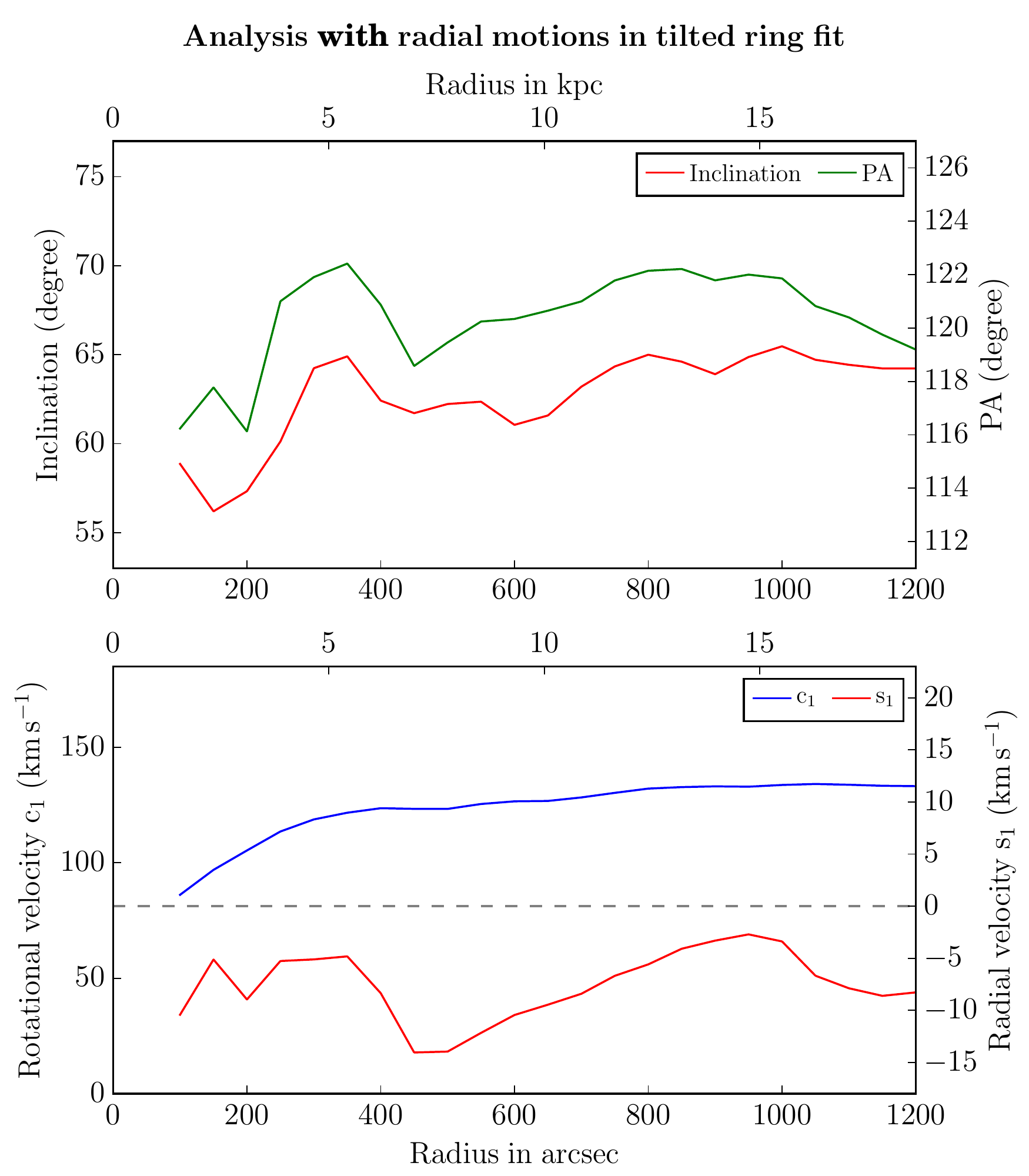}	
	\includegraphics[width=\widthdouble]{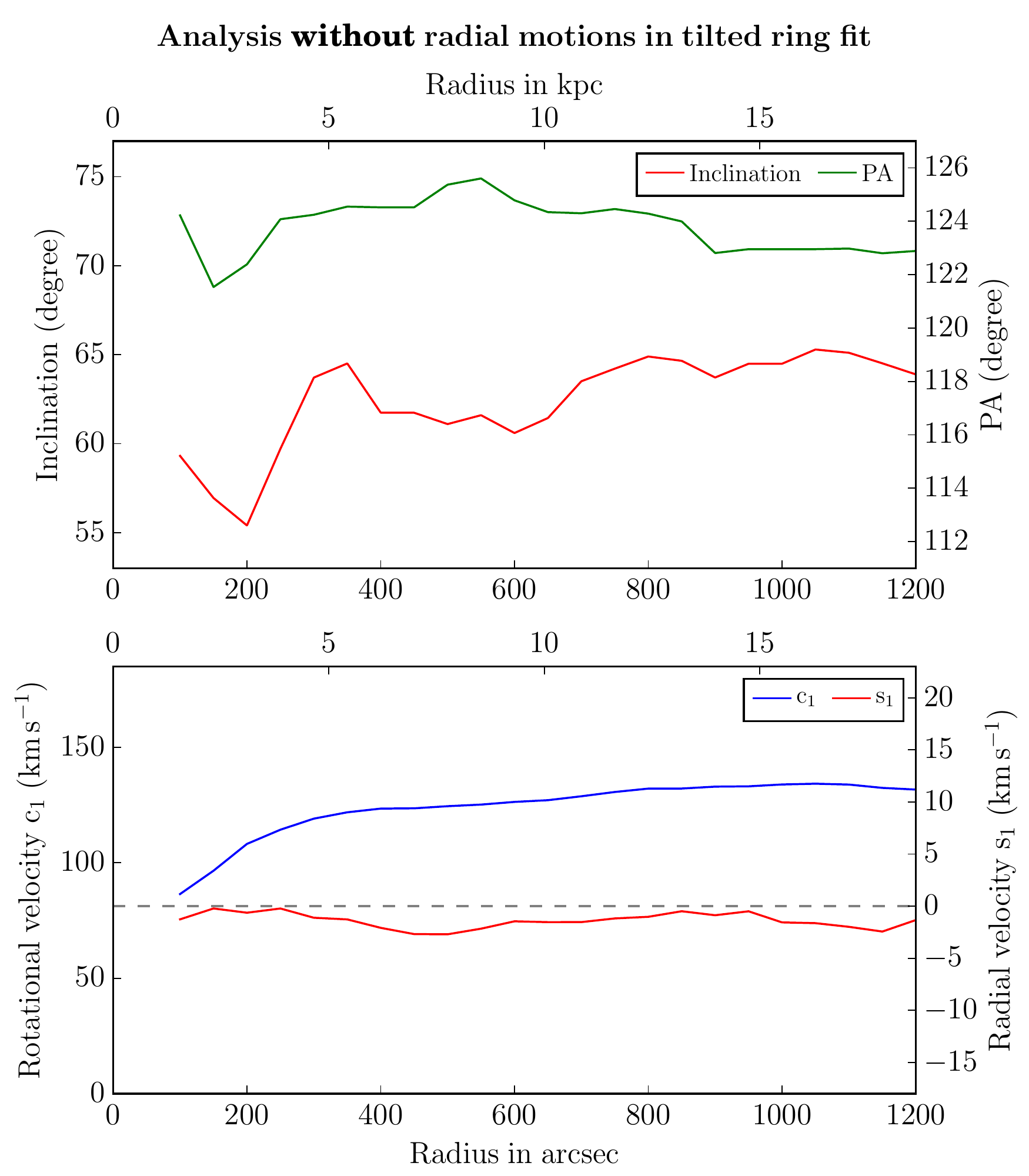}
 \end{center}
\caption{
Comparison of results obtained by two-step fits to the observed velocity field of NGC~2403, where first the ring geometry is determined in a tilted ring fit and subsequently a Fourier decomposition is performed.
Including a radial velocity term already in the tilted ring stage leads to the plots on the left, omitting such a term leads to the plots on the right.
The determined rotation curves and inclinations are similar but the position angle deviates by up to six degrees since in the second fit the geometry is optimised according to a model without radial motions. One sees that the position angle adjusts to compensate for radial inflow and thus the following Fourier decomposition is unable to recover a significant radial flow (red line in bottom right panel). If $i$, $PA$, $V_{\rmn{rot}}$ and $V_{\rmn{rad}}$ are fitted simultaneous during the tilted ring fit as in the left panels clear signs for radial motions are obvious (left bottom panel). Uncertainties are approximately $\pm 1\,\mathrm{km\,s^{-1}}$ for $s_1$ like in Figure~\ref{NGC2403}  
}
\label{NGC2403-noVrad}
\end{figure*}

The results of our comparison of the two schemes applied to the NGC~2403 data are presented in Figure~\ref{NGC2403-noVrad}. The derived rotation curves turn out to be nearly identical and also the inclinations are very similar. The measured position angle is systematically higher in the case where no radial motions are permitted in the tilted ring fit. The deviation in position angle of $6^{\circ}$ appears small and may be dismissed as not significant at first glance, however, it causes the measured $s_1$ component in the Fourier decomposition to change substantially.  If the ring geometry is determined without taking radial motions into account, the subsequent Fourier analysis is not able to recover any significant radial velocities in the second step of the analysis. The $s_1$ component varies between 0 and $-2.5\;\mathrm{km\:s}^{-1}$, which can be taken as a null result. This is in general agreement with the results of \cite{Schoenmakers1997} and \cite{Trachternach2008}, who also find little evidence for radial gas motions.

If in turn radial motions are already allowed in the tilted ring fit, the conclusions change and we find clear signs of radial inflow. The $s_1$ component is always well below zero and shows a large feature between radii of 350'' and 900'' with a maximum inflow velocity of $15\;\mathrm{km\:s}^{-1}$. In this scheme, the radial motions are already part of the tilted ring fit, which itself does not depend on any a-priori assumptions except the centre position. We conclude that taking the radial velocity term into account in the first analysis step makes this approach more general and allows us to recover a radial velocity flow if it is present in the data. 

To better quantify the differences between the two approaches, we create a synthetic velocity field, similar to the one derived from the NGC~2403 data. We choose the disk to be flat with a constant position angle of $122^{\circ}$ and an inclination of $60^{\circ}$. For the rotation velocity, we adopt a rotation curve of Brandt-type \citep{Brandt1960} in the rising part and a constant velocity of $135\;\mathrm{km\:s}^{-1}$ outside of 500''. To this ideal disk we add an inflow component starting at a radius of 350'' and linearly increasing to $15\;\mathrm{km\:s}^{-1}$ at 450''. It then declines again linearly and returns back to zero at 950''. The exact shape of the rotation curve is not crucial, since the inflow we try to recover happens only in the flat part of the rotation curve.

\begin{figure*}
 \begin{center}
	\includegraphics[width=\widthdouble]{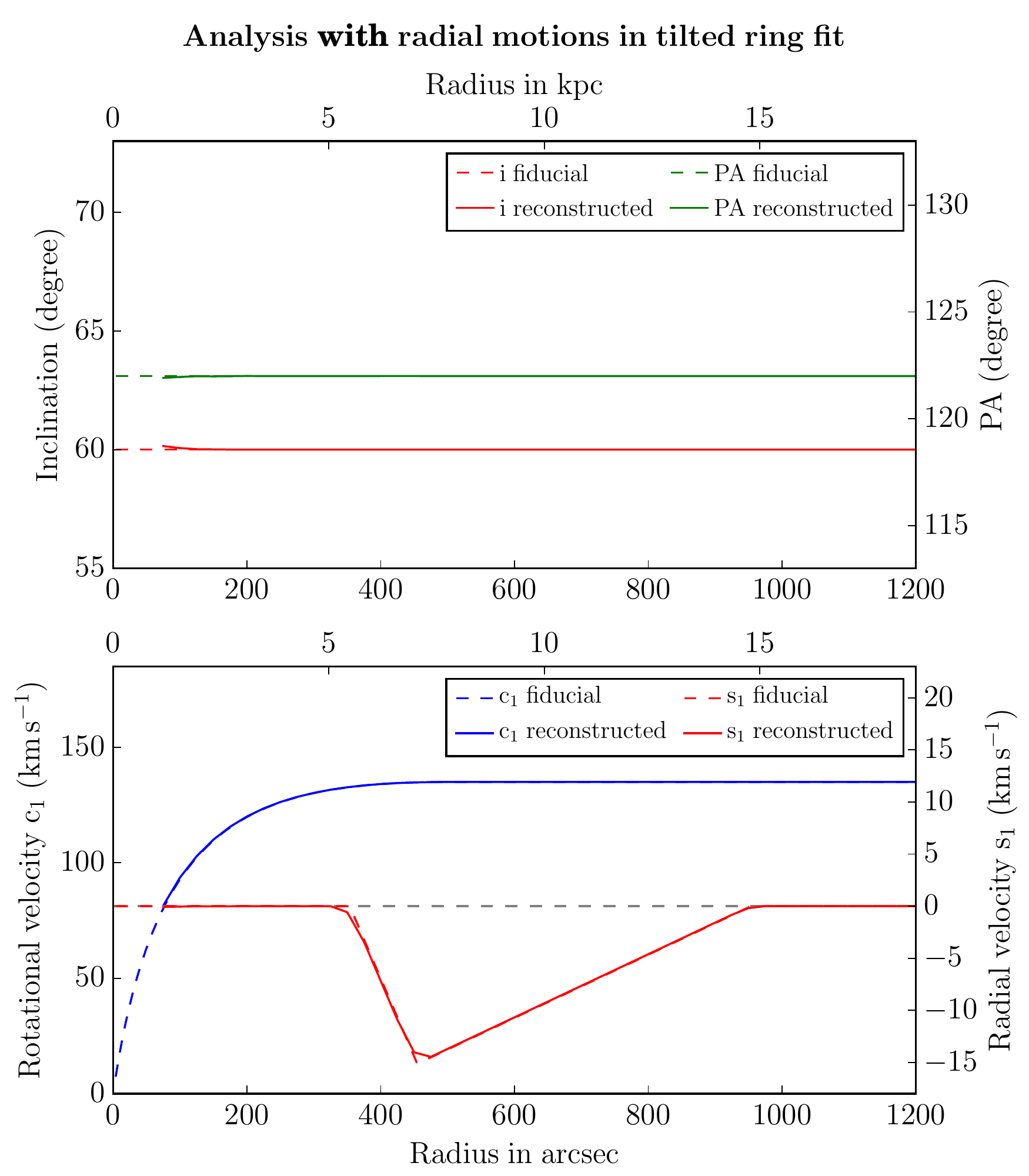}	
	\includegraphics[width=\widthdouble]{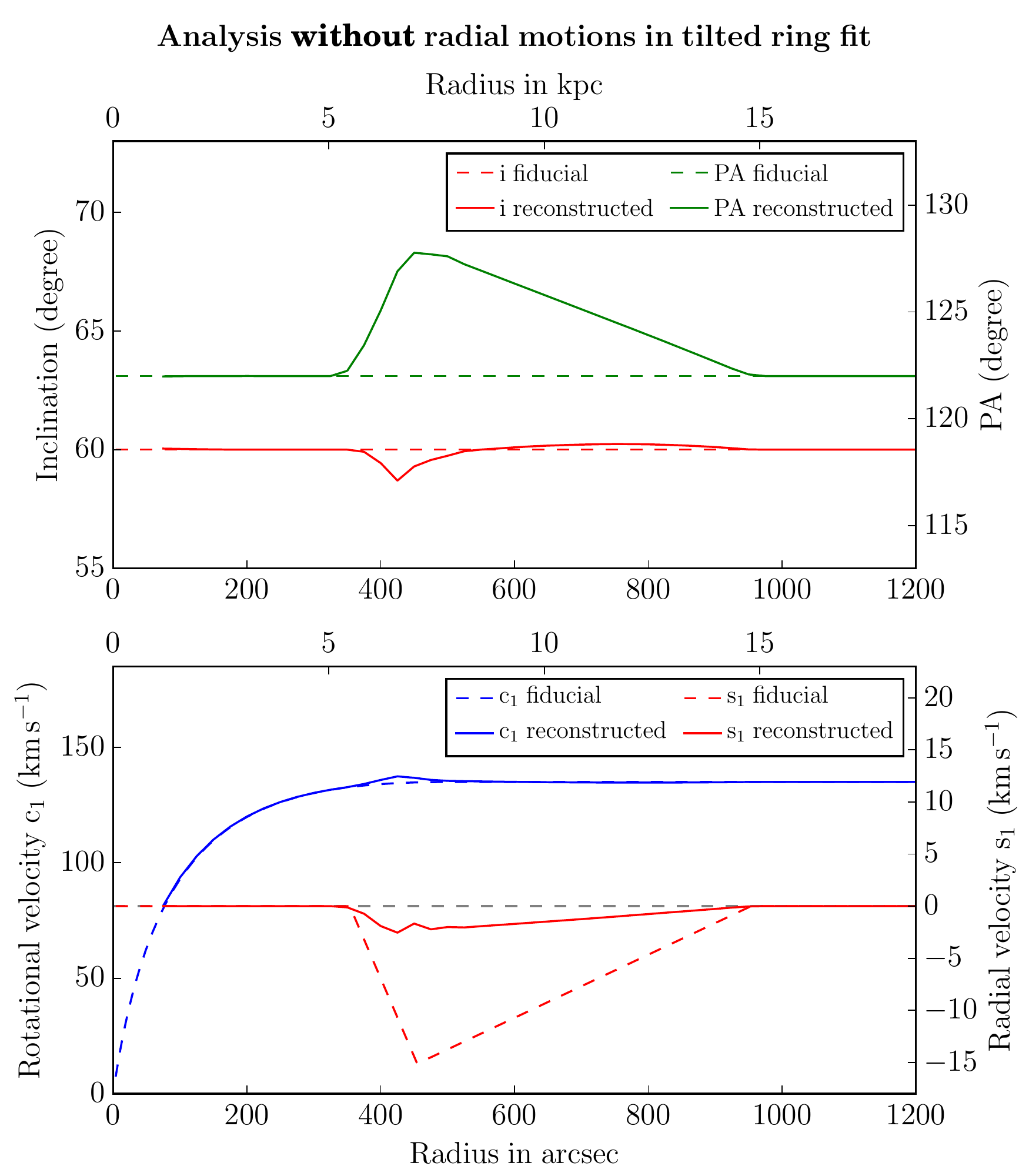}
 \end{center}
\caption{The two different two-step fitting procedures (with and without a radial velocity term in the tilted ring fit) applied to a synthetic velocity map that roughly resembles the one for NGC~2403. In each plot the input values used to create the artificial data are shown as thin dashed lines. 
The plots on the left hand side show the results derived by the fitting scheme which includes radial motions already in the tilted ring fit. The input values are exactly recovered, i.e.for $i$, $PA$ and $V_{\rmn{rot}}$ input values and fit results are indistinguishable. For the $s_1$ component the plot shows minor deviations. 
The results of the scheme in which the ring geometry is determined without taking radial motions into account are shown on the right side. During the initial tilted ring fit the position angle is fitted to a value that compensates for the radial motion (upper right panel). In the subsequent Fourier decomposition only a small fraction of the actual radial velocity signal is recovered. The Fourier decomposition is not able to find the correct inflow because it is restricted to the wrong ring geometry.}
\label{Mock2403}
\end{figure*}

We then fit this velocity field in the same way we treated the NGC~2403 data. We first apply a scheme that includes a radial component already in the tilted ring analysis, and then use the method that does not take it into account. Noise is not added to the mock data since our focus is solely on the systematics of both methods. In addition, simple Gaussian noise would largely average out due to the huge number of data points in the mock velocity map. The results are shown in Figure~\ref{Mock2403}.

The analysis with inflow in the tilted ring stage perfectly recovers all parameters. There are only small deviations of the $s_1$ component in the innermost parts of the disk and some finite-resolution effects. The solid curves representing the fit are indistinguishable from the dashed lines representing the input of the mock velocity field and most importantly, the fit recovers the correct inflow pattern.

In contrast to this, the analysis method that does not include radial motions in the tilted ring fit is not able to properly recover the radial flow. It shows some signal in the $s_1$ component, but the maximum amplitude of $\approx3\;$km~s$^{-1}$ is only a small fraction of the correct value. One can see that the position angle determined in the tilted ring analysis yields an incorrect value to compensate for the inflow. The deviation is roughly proportional to the radial velocity and peaks at a mismatch value of about $6^{\circ}$.  The overall picture is similar to the deviation between the two methods seen for the NGC~2403 data. Since the radial velocity component is already compensated for by the disk geometry in the tilted ring stage, the subsequent Fourier decomposition cannot recover more than a small residual $s_1$ component, and thus disguises potential signatures of substantial radial gas motions. 

\subsection{Simultaneous fit}
\label{SimultaneousFit}

As demonstrated above, it is necessary to include the radial velocity component already in determining the ring geometry. The next question is, what happens to the second or higher orders in the Fourier decomposition.  From our experience with fits to mock velocity fields, we come to the conclusion that the second order terms do not couple to $i$ and $PA$ in a significant way. It should therefore be no problem to fit them in a subsequent step, once  the ring geometry was determined in a tilted ring fit. However, since they are orthogonal not only to the first order terms, but also to $i$ and $PA$, it is also only a minor additional complication to fit them all simultaneous. To avoid any issues from carrying out a fit in two distinct steps, we investigate a fitting scheme where we fit inclination, position angle and all velocity components up to second order at the same time. Compared to the two-step approach including radial motions in the tilted ring analysis, the simultaneous fit yields nearly identical results. It is equally reliable and stable. 

Before outlining the details of the simultaneous fitting approach that we apply to the THINGS data in Section~\ref{Fit_Procedure}, we comment on the prospects of including terms that are higher than second order in the Fourier decomposition.  We find that this is a difficult exercise.  For example, the $c_3$ component is highly degenerate with the inclination. This makes a simultaneous fit of $c_3$ and $i$ impossible. One cannot compare the results of a two-step approach to a simultaneous fit of third order. A third order Fourier decomposition based on a disk geometry determined in a first or second order analysis cannot yield the correct value for the $c_3$ term. The parameter $c_3$ is implicitly assumed to be zero in the first step, and so the inclination is fitted to a value that compensates for a possible non-zero $c_3$ term.  One could in principle fix $c_3$ and fit everything else up to fourth order, but there is no justification for assuming the $c_3$ component to vanish or to be of any particular form while at the same time trying to measure the $s_3$ and higher components. Therefore it seems there is no satisfying way to safely deduce higher than second order Fourier parameters without the risk of systematic errors, at least not from the velocity field alone and in a purely kinematic approach. 
It is for this reason that we restrict our analysis to a simultaneous fit of second order in $V_{\rmn{los}}$. This limits the level of detail we can derive but on the other hand reduces the necessary assumptions and related systematics to an absolute minimum while also 
making the model as flexible as possible since everything is fitted simultaneous.

\subsection{Fit procedure}
\label{Fit_Procedure}

The model we finally adopt is a second order Fourier decomposition of the following form:
\begin{equation}
\label{Eq9}
 \begin{aligned}
 		V_{\rmn{los}}		& =	V_{\rmn{sys}} + \sin(i) \, \Big[	\, c_0	& + s_1 \sin(  \theta) \; \;	 & + c_1 \cos(  \theta) 							\\
 		 								&																						& + s_2 \sin(2 \theta)		 		 & + c_2 \cos(2 \theta) 	\; \Big]\;.		
 \end{aligned}
\end{equation}
The coordinates of the disk frame ($R$,~$\theta$) are related to the sky coordinates ($X$,~$Y$) by equations~\ref{trafo1} and \ref{trafo2}. The coordinate transformation itself depends on the galaxy centre, the inclination and the position angle.  We fit all five Fourier parameters ($c_0$, $s_1$, $c_1$, $s_2$, $c_2$) as well as $i$ and $PA$ for each ring simultaneously. The centre coordinates ($X_C$,~$Y_C$) and the systemic velocity have to be determined in advance and are kept fixed for all rings. We do that by fitting a simple tilted ring model in which $V_{\rmn{sys}}$ and the centre coordinates are free to vary to the innermost area of the velocity field. In this part of the disk, the fit is stable even without assuming a fixed centre. We then take the average values of the ring fits. The choice of $V_{\rmn{sys}}$ only influences the $c_0$ component in the Fourier decomposition. All other parameters are unaffected by that and an incorrectly chosen $V_{\rmn{sys}}$ remains transparent throughout the analysis.

The ring widths are chosen for each galaxy individually to get the best compromise between good sampling and fit stability. Typical widths are between 24'' and 48'' which is always larger than the THINGS synthesised beam width of $\approx11$" \citep{walter2008}.
In the common case of warped disks, gaps naturally appear between rings and the observational data within the gaps of our tilted ring model are unused. To alleviate this effect to some degree and to make maximal use of all available data, we choose the separation of adjacent rings to be 37\% smaller than the width of the rings. This closes most of the gaps in the model but also leads to some overlap of rings. The exact amount of overlap depends on the fitted ring geometry. Since each ring is fitted individually and independently of all other rings, the overlap has no influence on the fit result. The only consequence is that neighbouring rings are statistically slightly correlated (however, omitting every second ring in our analysis and the following figures leads to a statistically independent sample). 

We start the fit procedure with the central ring. After the fit for this ring has converged, the fit results are taken as initial values for the second ring. This procedure of outward forwarded initial values is used for all succeeding rings. We only have to manually set the initial values for the central ring to reasonable starting values. Their exact choice has negligible impact on the overall result.
In general, the question of whether the fit for a particular ring converges or not depends more strongly on the quality of the data and on possible irregularities in the velocity field than on the given initial conditions. For good quality data, the fit converges to the right value even if extremely poor guesses for the initial conditions are made. For disturbed disks, low inclinations, and especially velocity fields with linearly rising rotation curves, the rings have the tendency to flip to a very high inclination. This often happens in the innermost parts of the galaxies, and is likely caused by solid body rotation and bars. If this affects only the few innermost rings, we can simply ignore them and use the well fitted outer part, since this is where our scientific interest is focused on. 

Linearly rising rotation curves make it impossible to disentangle i and $V_\rmn{rot}$. It can especially in the centre of galaxies happen that the fit converges to obviously incorrect values (i very close to 90). If this affects only a few rings in the galaxy centre while the other rings are fine, we discard the obviously wrong ones and keep the rest.

After the ring fit is completed, we calculate the $V_\rmn{rad}$ and $V_\rmn{rot}$ parameters according to Equation~\ref{Eq8}. As mentioned before, the degeneracy between $B_1$ and $C_1$ cannot be broken and we assume $C_1 = B_1$.

The final step is to convert velocities to mass flow rates. From the derived Fourier parameters, we construct a two-dimensional synthetic radial velocity field by taking only Fourier components of radial direction into account. 
To convert the THINGS H\,{\sc i} integrated line intensity (zeroth moment) maps to column densities and finally to H\,{\sc i} mass surface density, we adopt the method and distances given in \citet{walter2008} and assume the H\,{\sc i} gas to be optically thin.
The THINGS first-moment maps represent an intensity-weighted and therefore approximately mass-weighted mean velocity at each position $(R, \phi)$. As outlined above, we decompose these velocities into a rotational, $V_{\mathrm{rot}}(R,\phi)$, and radial, $V_{\mathrm{rad}}(R,\phi)$, component. In order to compute a 2D distribution of the radial mass flow rate, we simply need to multiply $V_{\mathrm{rad}}(R,\phi)\times\Sigma_{\mathrm{HI}}(R, \phi)$ for each position. To obtain the net flow rate at each radius, we integrate over the azimuthal angle $\phi$: 
\begin{equation}
\label{Eq10}
\Gamma_{\mathrm{rad}}(R) = \int_0^{2\pi} V_{\mathrm{rad}}(R,\phi) \; \Sigma_{\mathrm{HI}}(R, \phi) \; R \: d\phi \;.
\end{equation}

Here, $\Gamma_{\mathrm{rad}}(R)$ denotes the radial mass flow rate and $\Sigma_{\mathrm{HI}}$ the H\,{\sc i} mass surface density. We note that radial velocities can be directed inward or outward so that $\Gamma_{\mathrm{rad}}(R)$ represents a net flow rate. We also note that in practice we calculate $\Gamma_{\mathrm{rad}}(R)$ in radial bins of finite width $\Delta R=24-48''$ (compare description earlier in this section). We interpret these radial H\,{\sc i} net mass flow rate profiles and compare them to SFR profiles of our galaxy sample in Section~\ref{continuity}.

\subsection{Error estimation}
\label{ErrorEstimate}

For the fit procedure described in Section~\ref{Fit_Procedure} we use a standard $\chi^2$-optimiser, which also gives formal statistical uncertainties. However, in order to obtain more realistic uncertainty estimates, we carry out additional Monte-Carlo modelling. That is, we add random noise with specific characteristics to our data and measure how this influences the results. We repeat this 100 times to build up a statistically significant sample. 

We point out that adding purely Gaussian noise would not serve our purpose, because it would average out due to the large number of data points in the H\,{\sc i} maps. We are more concerned about potential systematic patterns in the velocity field that we cannot account for in our fitting routine. \citet{Sellwood2010} advertise a concept that uses the residuals of the fit as a noise source. The idea is that every residual structure not included in the model is in fact noise to the fit, and the residuals can be used as a prior to create a noise field with statistical characteristics that are well matched to the data. The authors randomise and rescramble the two-dimensional map of the residual by rotating and shifting chunks of it in radius by random amounts. This creates a statistical noise field, but it also preserves any structures and coherence on small scales. 

We follow the same approach, but we modify the rescrambling procedure since in our model inclination and position angle are allowed to vary throughout the disk. We therefore do not deal with one continuous velocity field but instead with an ensemble of tilted rings that can in general have gaps and overlapping regions. We therefore apply this procedure to each ring individually. The exact procedure we use is the following: After the fit, we calculate for each ring the residual after subtracting the fit from the data. Each residual is deprojected according to the ring geometry determined in the fit, and it is rotated around the ring centre by an individual angle, randomly chosen between $-\pi$ and $\pi$. It is then reprojected using the same geometry. An artificial noise map is created by combining all rings, which is then added to the observed velocity field. The perturbed data cube is then fitted again. The resulting residuals are rescrambled in the same way and added to the original data to be used in the next iteration. We repeat this procedure 100  times, and so obtain an ensemble of 100 values for our fit parameters. From this, we compute mean and standard deviation which we take as our final results. We investigated the number of iterations necessary for convergence by running up to 2500 iterations and found that 100 is sufficient.

\subsection{Comparing radial H\,{\sc i} mass fluxes to star formation rates}
\label{continuity}

The processes that transport gas within a galaxy and transform it into stars are complex. In order to compare our H\,{\sc i} radial flux measurements to the SFR profiles, we present a simplified model that is solely based on mass continuity arguments:
\begin{equation}
 \frac{d}{dt} \int_A \Sigma_\mathrm{HI} \; d\mathrm{A}= - \oint_{\delta A} \vec{\F}_\mathrm{HI} \; \vec{dl} \;. 
\end{equation}
$\Sigma_\mathrm{HI}$ is the H\,{\sc i} column density or equivalent the H\,{\sc i} mass surface density and $\vec{\F}_\mathrm{HI}$ the H\,{\sc i} mass flux density. We assume the disk to be thin and therefore only deal with a two-dimensional problem.
If we include the star formation rate surface density $\Sigma_\mathrm{SFR}$ as a sink term on the left hand side, focus only on the radial mass flux and use Gauss's theorem to express everything in local quantities, we find:
\begin{equation}
 \frac{d}{dt} \Sigma_\mathrm{HI} - \Sigma_\mathrm{SFR} = - \frac{d}{dr} \F_\mathrm{rad} \;.
\end{equation}
For this study we only deal with azimuthally averaged quantities. This justifies ignoring rotational streaming in the previous step and leads to the following relation: 
\begin{equation}
 \label{pile-up}
 \int_0^{2\pi} \left( \frac{d}{dt} \; \Sigma_\mathrm{HI} - \Sigma_\mathrm{SFR} \right) \; r \: d\phi = - \frac{d}{dr} \; \int_0^{2\pi} \F_\mathrm{rad} \; r  \: d\phi \;.
\end{equation}
The quantity $\Gamma_\mathrm{rad}= \int_0^{2\pi} \F_\mathrm{rad} \; r  \: d\phi$ is our measured radial mass flux profile.

We note that we do not take into account the contribution from helium when we quote H\,{\sc i} masses or surface densities (a factor of 1.36). For an appropriate description, one probably also has to add additional sink and source terms to the left-hand side. These could for example reflect stellar winds and outflows, galactic fountains or gas that condenses from the hot halo gas or circumgalactic medium and settles to the disk. These processes are in general difficult to model and beyond the scope of this toy model. However, if we measure a violation of Equation~\ref{pile-up}, we have an indication that some of these processes may be present. 

If one assumes the special case that star formation and other sink and source terms vanish and observes the radial mass flux 
to be constant (but not necessarily zero) over some radial range, then the H\,{\sc i} column density should not change with time and the galaxy would be in a (quasi) stationary state. On the other hand, if $\frac{d}{dr} \Gamma_\mathrm{rad}$ is negative, gas should pile-up.

If we assume our galaxies to be in such a stationary state and therefore $\frac{d}{dt} \: \Sigma_\mathrm{HI} = 0$, the mass flux at a given radius equals the total star formation within this radius:
\begin{equation}
 \label{mass_conservation}
 \int_{0}^{R} \int_{0}^{2\pi} \Sigma_\mathrm{SFR}(r, \phi) \; r \: d\phi \; dr =  - \int_{0}^{2\pi} \F_\mathrm{rad}(R, \phi) \; r \: d\phi \;.
\end{equation}
This is a relation we can test by comparing our radial mass flux measurement to cumulative star formation rate profiles, which we derive from GALEX FUV data. For this we closely follow the procedure described in \citet{Bigiel2010a} and use the conversion factor from \citet{Salim2007}. We take the ring geometry from our kinematic analysis and for each ring compute the total SFR from the galaxy centre out to this radius. We correct for Galactic extinction according to \citet{Schlafly2011} and blank bright stars but neglect extinction within the target galaxies. 
A detailed comparison of the various SFR calibrations using different tracers (e.g., H$\alpha$, UV or IR emission) is beyond the scope of this study. To be consistent with accurate, published galaxy integrated SFRs, we scale our cumulative SFR profiles to match the literature values (Table~\ref{table1}) at large radii, which is equivalent to assuming a constant intrinsic extinction. We note that this approximation is sufficient for the purposes of this study and any more accurate treatment of internal extinction at small radii does not affect our conclusions.

As we show in Section~\ref{Results}, the naive assumption of vanishing time dependence and absence of other processes is rarely appropriate, but the deviations from this simplified model might give insights on the magnitude of other processes.

\section{Verification of the method}
\label{Verification}

As we show in Section~\ref{TwoStepFit}, the overall fitting scheme can be in general quite sensitive to prior assumptions. We therefore conduct a suite of tests in order to assess the correctness of our results and to determine the potential limitations of our approach. We focus mostly on the analysis of synthetic velocity fields and on testing under which conditions the input values can be recovered correctly. But we also investigate the dependence of our results on the type of velocity maps (such as intensity-weighted first-moment or Hermite) that are used.

\subsection{Limits of the method}

In Figure~\ref{Mock2403} we demonstrate that a two-step fit with radial components at the tilted ring stage is perfectly able to recover radial motions from a synthetic velocity field, at least if the disk is flat. This is also true for the second order simultaneous fit scheme introduced in Section~\ref{Fit_Procedure} which is used for the rest of the paper. However, galactic disks are not flat, but instead they exhibit warps and changes of position angle. Both constitute a great challenge for any tilted ring like analysis, because the deprojection of the velocity field is no longer continuous and unique.

\begin{figure}
 \begin{center}
	\includegraphics[width=\widthsingle]{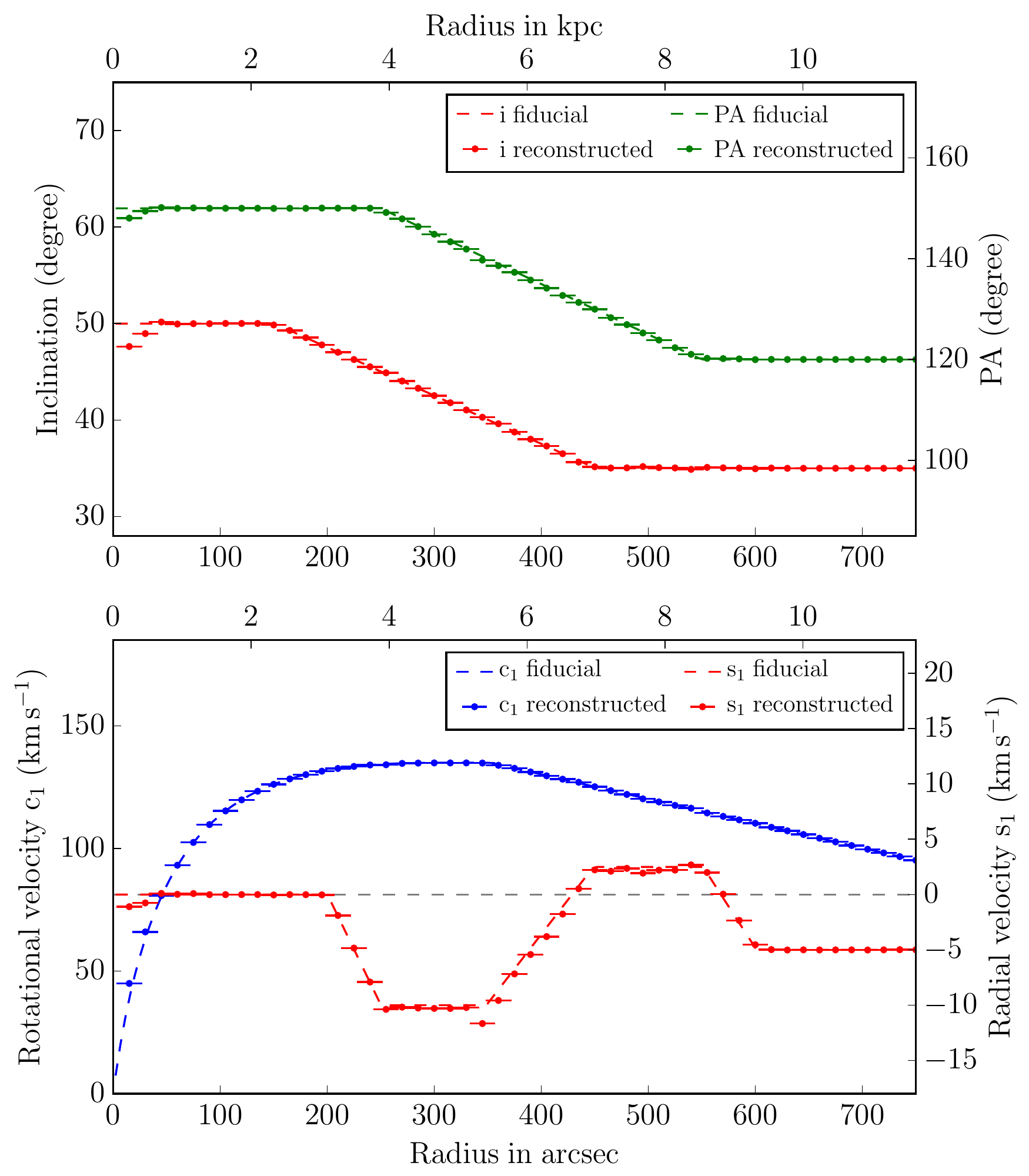}
	\caption{Our method applied to a synthetic velocity field. 
	Even if geometry and velocity parameters of the mock data deviate in different patterns and at different, overlapping radial ranges from a flat disk, the analysis is able to recover all of these features with high precision.}
	\label{fig:Mock03}
 \end{center}
\end{figure}

In Figure~\ref{fig:Mock03} we show an example of a fit to a synthetic velocity field with mild warps and position angle changes. We start again with a flat disk with a Brandt-type rotation curve. It is then modified by a decrease in inclination $i$ by $15^{\circ}$ over 300'' in radius, a change of position angle $PA$ by $30^{\circ}$ over 300'', and a decrease of rotation velocity of $10\;\mathrm{km\:s}^{-1}$ per 100'' starting at 350''. In addition, a pattern of in- and outflows of different magnitude up to $10\;\mathrm{km\:s}^{-1}$ is added. All these single effects cover different radial ranges. Therefore, a variety of combinations of effects is simulated at the same time. Again, since we are interested in systematic effects, no noise is added to the artificial velocity field.

The plots in Figure~\ref{fig:Mock03} show that our method can fully recover the parameters of the input velocity field to a very high accuracy. Problems only occur, as expected, at very small radii. In addition, a slight tendency for deviations exists at points where the slope of parameters abruptly changes. But overall, the result is highly satisfactory. So, our method is valid at least for mildly disturbed disks.

In the following we will address how our fitting procedure reacts to stronger warps. Generally, a decrease of inclination with radius will not be a problem, but a strong increase will be:

Taking two annuli with radii $R_1$, $R_2$, $R_1<R_2$ and inclination angles $i_1$, $i_2$, $i_1<i_2$, their apparent semiminor axes have the projected length $R \times \cos( i)$. 
If $R_1 \times \cos( i_1) = R_2 \times \cos( i_2)$, the two annuli completely overlap on the minor axis and our assumption of a thin disk with unique line-of-sight velocities is no longer valid. 
This condition can be converted to an expression for the maximum inclination increase that does not lead to a pile-up of rings on the minor axis:
\begin{equation}
	\left( \frac{\Delta i}{\Delta R} \right) _{crit} = \frac{1}{R \, \tan{i}} \;.
	\label{i_crit}
\end{equation}
Here, $\Delta R$ denotes the increase in radius from one ring to the next and $\Delta i$ the increase in inclination between the two rings, $i$ and $R$ are inclination and radius of the inner of the two rings. 

While limited overlap of adjacent rings is in practice uncritical (and even enforced by the rings chosen 60\% wider), an inclination increase exceeding the critical rate will make even non-adjacent rings overlap. The apparent minor axis of the rings shrinks with increasing radius. In such conditions, the tilted ring approach breaks down. 

Similar overlapping effect can also appear for changes of the position angle. Since the geometry of position angle changes is more complex, we do not derive an analytic expression for this case. In the end, it would anyway be necessary to consider the combined effect of inclination and position angle changes, which complicates things even further.

\begin{figure}
 \begin{center}
	\includegraphics[width=\widthsingle]{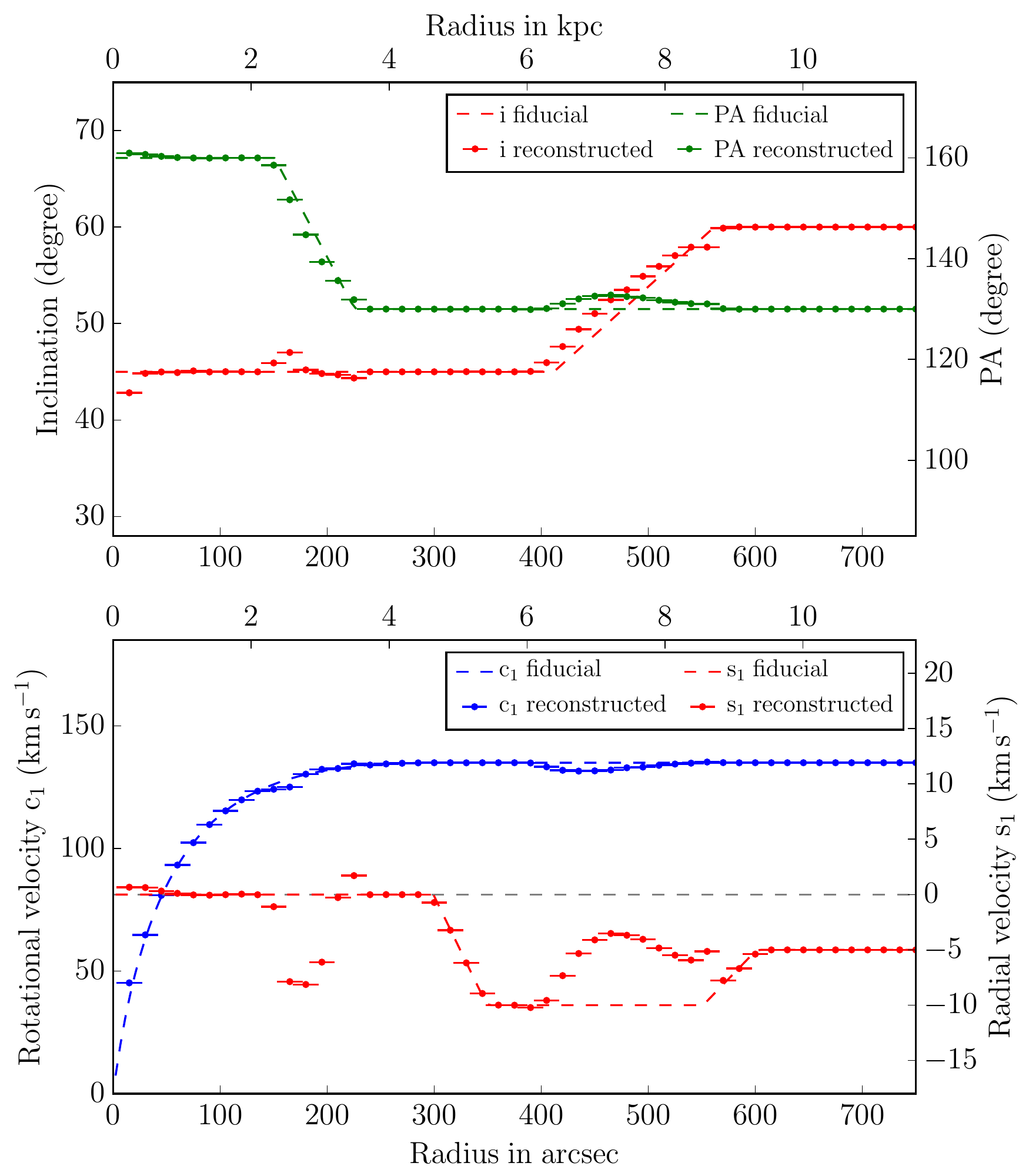}
	\caption{Fit to a velocity field that shows much stronger changes in position angle and inclination than the one used in Figure~\ref{fig:Mock03}. Obviously, the analysis is not capable anymore to fit these data correctly. The steep increase of $i$ with nearly twice the critical rate derived in Equation~\ref{i_crit} causes a mismatch of all parameters. The errors in $i$, $PA$ and $V_{\rmn{rot}}$ seem small but they result in a significant error in the $s_1$ term of up to $7\;\mathrm{km\:s}^{-1}$. Also, the change in $PA$ at $R\approx200''$ is now too fast to still allow a correct derivation of the $s_1$ term.} 
	\label{fig:Mock04}
 \end{center}
\end{figure}

Our experiences with fits to synthetic velocity fields are in good agreement with the above analytic considerations. In cases where the inclination decreases (e.g. Fig.~\ref{fig:Mock03}) or increases with less than the critical rate given in Equation~\ref{i_crit}, we see no significant deviations. If the inclination is increased close to the critical rate, some noticeable deviations appear, but the derived results are still usable with less than $1.5\;\mathrm{km\:s}^{-1}$ maximum error for the $s_1$ component (not shown). However, much larger changes of the inclination angle cause strong artefacts in the fit results.  This is illustrated in Figure~\ref{fig:Mock04}, where the inclination angle changes with twice the critical rate.  At the warp, the inclination is typically fitted to a too high value. This is because data at larger radii interfere with those from smaller radii. In addition, also the position angle and the rotation velocity are fitted to a slight mismatch. At first glance, these deviations do not appear severe, however, as discussed earlier, they lead to  large deviations for the important $s_1$ parameter, with errors of of $5\;\mathrm{km\:s}^{-1}$ and more over a large range in radius. This is of the same order of magnitude as potential inflow velocities, and  therefore can render the fit result useless. Interestingly, no deviation for the radial component is seen if the $s_1$ component in the synthetic velocity field  is set to zero, although the derived values for $i$ and $c_1$ actually differ from the input. 

Also visible in Figure~\ref{fig:Mock04} is an artefact in the $s_1$ curve due to a rapid change of the position angle, which is not perfectly accounted for. The inclination and rotation curve stay quite stable with only minor deviations, but the radial velocity shows incorrect peaks of up to $10\;\mathrm{km\:s}^{-1}$. The $30^{\circ}$ change in $PA$ over only 75" in radius is a very fast one. If this change is stretched over 150'' we find only small deviations of less than $1\;\mathrm{km\:s}^{-1}$. At a rate of $30^{\circ}$ over 300'' as in Figure~\ref{fig:Mock03}, any significant deviations have vanished. 

We also investigate the influence of an incorrect determination of the galaxy centre. Our tests showed that an offset of the order of a few arcseconds away from the true centre is not critical for our analysis. It creates artefacts, but their amplitude quickly decrease towards larger radii. For example, offsets of the order of 5'' and 10'' are no longer noticeable outside of 100'' and 200'', respectively. We conclude that the uncertainties involved in determining the galaxy centre are therefore not an issue within the context of this study.

\begin{figure}
 \begin{center}
	\includegraphics[width=\widthsingle]{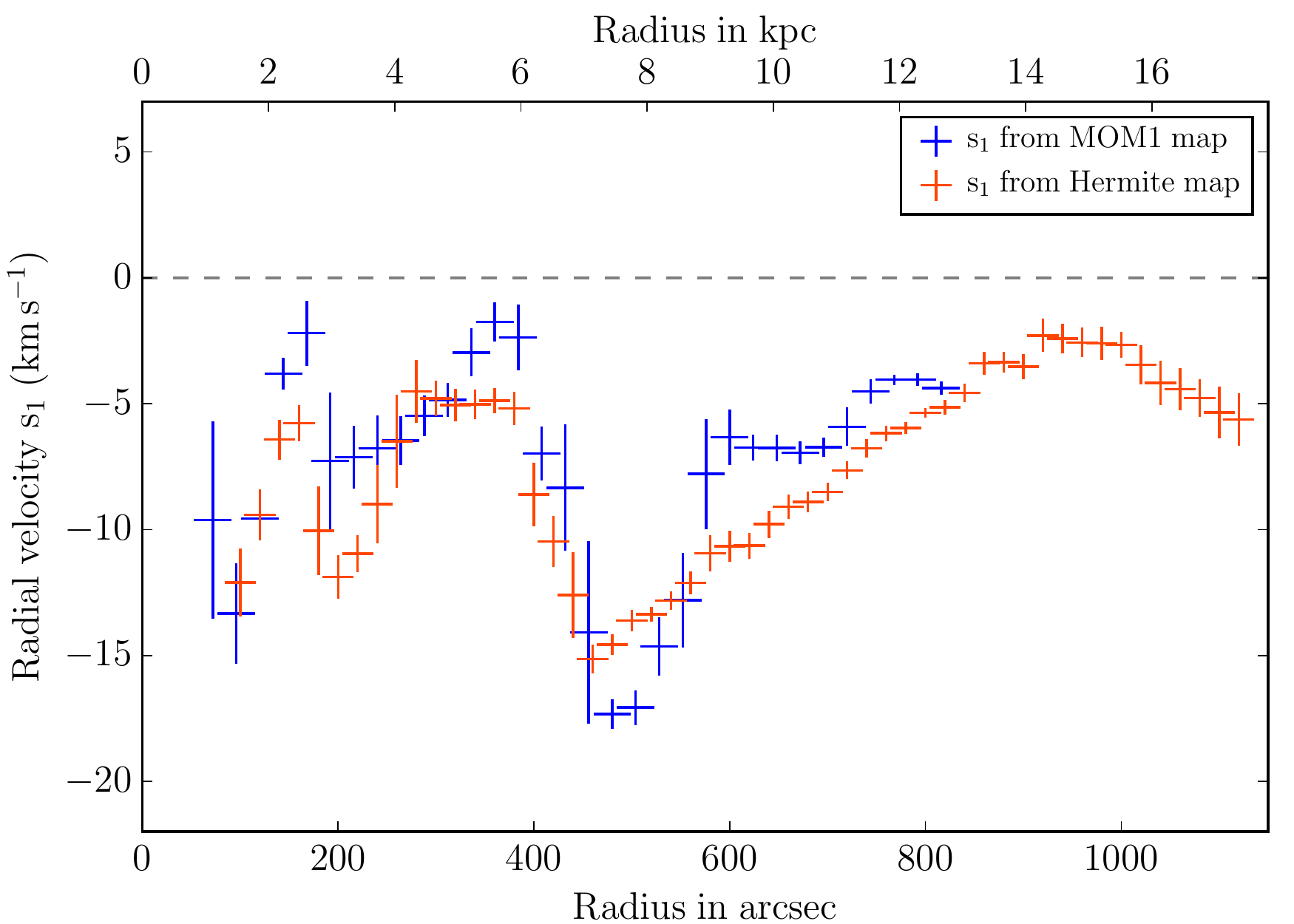}
	\caption{A comparison of the radial inflow pattern derived from a standard THINGS first moment map and from a velocity map computed by fitting Hermite functions to the H\,{\sc i} spectra. The Hermite map gives slightly different results than the moment map: From 450'' to 550'' the inflow rate is higher, from 600'' to 800'' lower. However, the overall distribution of radial velocities is quite comparable. We note that the moment map extends to further radii, which is important for the analysis presented here. 
	} 
	\label{NGC2403_Hermite}
 \end{center}
\end{figure}

\subsection{Comparison of first moment and Hermite velocity maps}

First moment maps are often employed to derive velocity maps from data cubes. They are easy to compute and do not require very high S/N levels. However, if the line profile is asymmetric or has extended wings, this method does not properly represent the bulk velocity of the gas. We therefore also include velocity maps from  \citet{deBlok2008} in our analysis. These are based on fitting Gauss--Hermite functions to the H\,{\sc i} spectra of each spatial resolution element. By modifying the Gauss function by the third order Hermite polynomial, the fitting function is capable of adapting to asymmetric line shapes and in this case better represents the bulk velocity. A detailed description can be found in \citet{vanderMarel1993}.

We apply our method to the galaxy NGC~2403 and compare the standard moment map provided by THINGS to the  corresponding Hermite map. The result is illustrated in Figure~\ref{NGC2403_Hermite}. Both velocity maps do not give exactly the same result but they are in general agreement. The most obvious difference lies in the intermediate part of the galaxy. In the moment map we find a maximum inflow velocity of $15\;\mathrm{km\:s}^{-1}$ at 450''. The inflow then linearly decreases to $3\;\mathrm{km\:s}^{-1}$ at 900''. The Hermite map shows even more inflow at around 500'' followed by a rapid decrease to a roughly constant inflow velocity of $6\;\mathrm{km\:s}^{-1}$ between 600'' and 800''. Beside this, the results are quite similar and most importantly, both maps show the same unambiguous signs of large scale inflow. 

Unfortunately, fitting  Hermite functions to the data cube requires substantially higher S/N compared to the moment maps. It can therefore not be applied to the outermost parts of the H\,{\sc i} disk. Based on the results of this comparison we decide to proceed using the moment maps for our analysis. The derived values for Hermite- and moment maps approximately agree with each other and the moment maps extend further out from the galaxy centres which is  particularly important for our aim of detecting large-scale radial inflow.

\section{Results \& Discussion}
\label{Results}

In the following section we present our results. For each galaxy we present four plots describing the fit parameters as functions of radius. Radius and velocities are measured within the galactic plane. 
For the exact details of the analysis see Section~\ref{Fit_Procedure}. The four plots contain the following information:
\begin{enumerate}
\itemsep 0em
\item inclination and position angle,
\item the Fourier parameters $c_1$ and $s_1$, which directly correspond to the principal components of the rotation and  radial velocity, respectively, 
\item the Fourier parameters $c_0$, $c_2$ and $s_2$, which give information about the azimuthal dependence of the velocity field,
\item the calculated radial H\,{\sc i} mass flux, not accounting for helium, $\mathrm{H_2}$ or other gas fractions and the cumulative star formation rate profile giving the total SFR within a certain radius based on GALEX FUV images and scaled to the more precise literature values given in Table~\ref{table1}. 
\end{enumerate}

The rotation direction of the galaxies is determined from optical images, especially from the more pronounced appearance of dark filaments on the nearside of the disk. By combining this information with the velocity field, one can distinguish if the galaxy is rotating clockwise or counter-clockwise, which also determines if potential radial motion is directed inwards or outwards.

\subsection{NGC 2403}

\begin{figure}
 \begin{center}
	\includegraphics[width=1.04\widthbig]{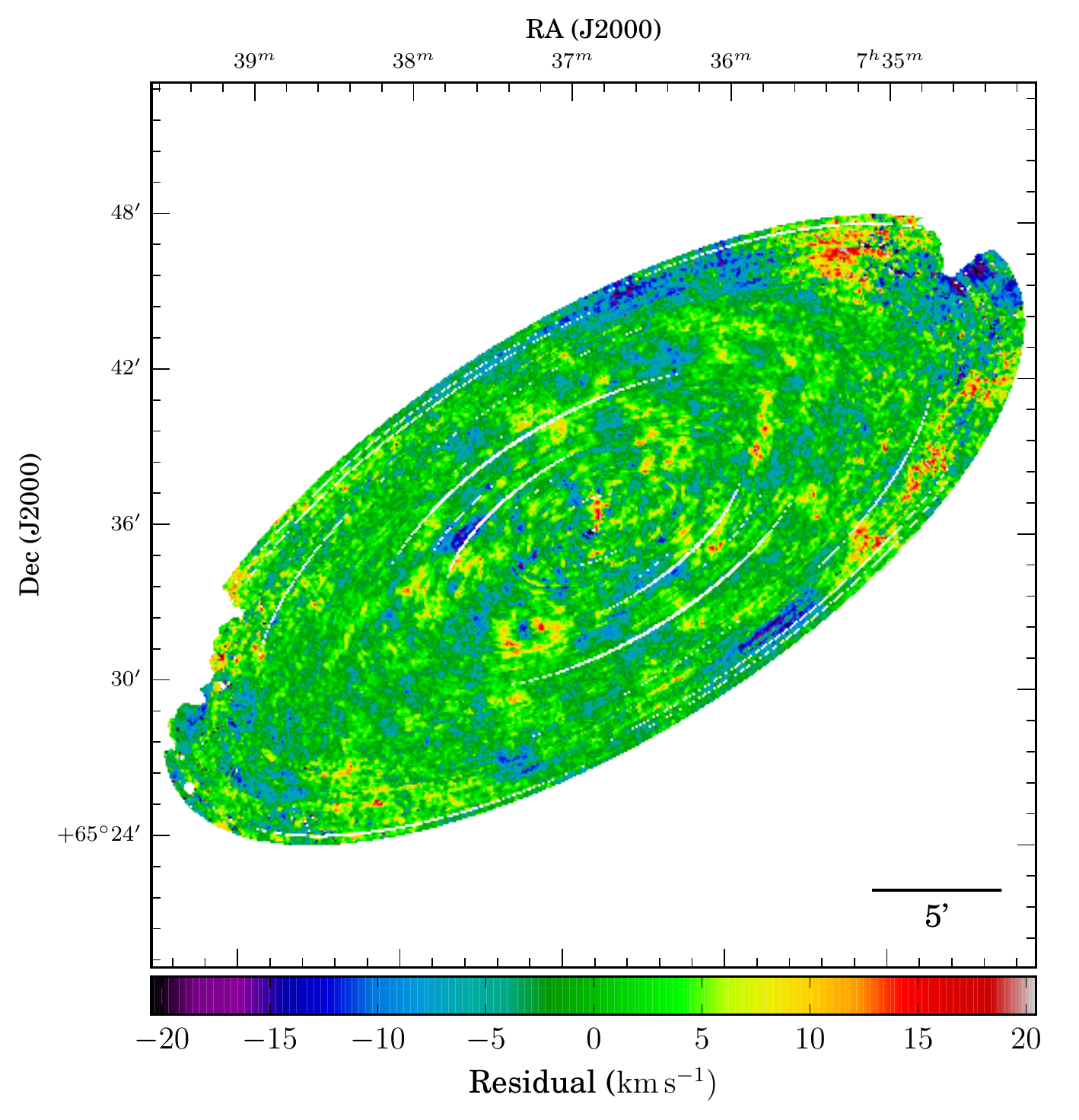}
 \end{center}
\caption{Residuals of the fit for NGC~2403. No distinct pattern of large-scale structure appears. A slight signature of spiral arms might be  visible in the inner part.} 
\label{NGC2403_Residuals}
\end{figure}

\begin{figure*}
 \begin{center}
	\includegraphics[width=\linewidth]{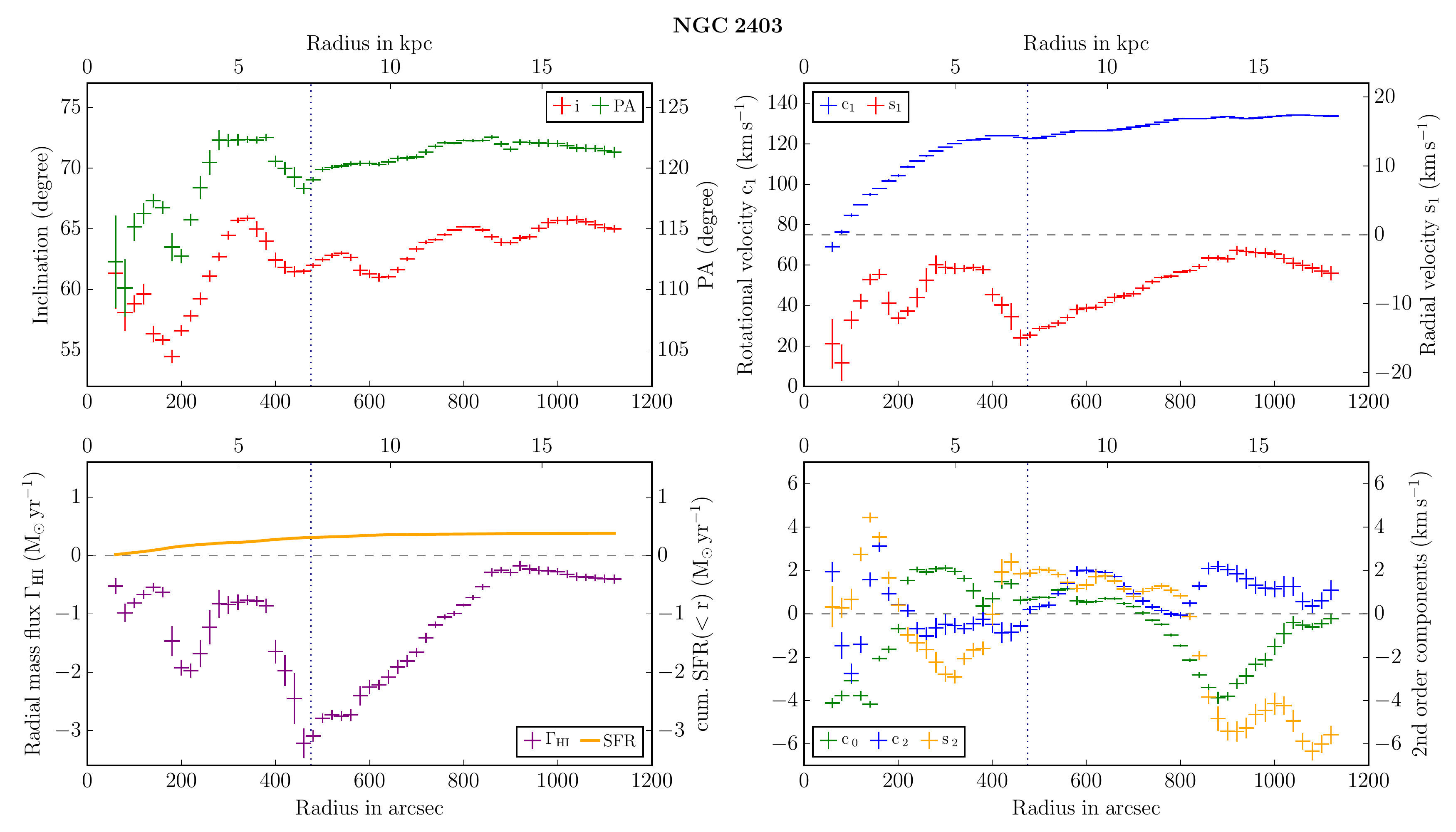}
	\caption{
	Disk geometry and Fourier parameters for NGC 2403. Horizontal error bars denote the ring width, vertical bars represent the errors estimated according to the scheme described in Section~\ref{ErrorEstimate}. The dotted vertical line indicates the optical radius $r_{25}$. The galaxy shows a smooth and unperturbed disk. Position angle and inclination only vary within limited ranges (top left panel). The radial velocity shows a very distinct signature between 400'' and 900'' (top right panel, red points) with a peak inflow velocity of $15\;\mathrm{km\:s}^{-1}$ which corresponds to am H\,{\sc i} mass flow of up to 3~M$_{\odot}\,\mathrm{yr}^{-1}$ (bottom left panel). The inflow quickly ceases inside of 400'' and shows a second smaller feature at 200''. For very large radii we find inflows of about 0.5~M$_{\odot}\,\mathrm{yr}^{-1}$. The bottom left panel also gives the cumulative star formation rate. For NGC~2403 basically all star formation takes place within the central 7~kpc.
	The bottom right panel displays the zeroth and second order terms which indicate the deviation from rotational symmetry.
	}
	\label{NGC2403}
 \end{center}
\end{figure*}

In our sample of 10 THINGS galaxies, NGC~2403 turns out to be the most suitable target for detecting radial mass inflow patterns. It has a large apparent size of roughly 16' x 10', was observed with good signal-to-noise ratio, and most importantly, it has a very flat and smooth H\,{\sc i} disk and regular velocity field. Our fitting results are summarised in Figure~\ref{NGC2403}. 

Position angle $PA$ and inclination $i$ are plotted as function of radius in the top left part of the Figure. The galaxy is rotating clockwise, and so the south-western part of the disk is closer to Earth. The rotational velocity (Fourier component $c_1$) is between 120 and 140$\;\mathrm{km\:s}^{-1}$. As already described in Section~\ref{TwoStepFit}, we find clear evidence for radial inflows.  The derived radial velocity (Fourier component $s_1$) is always negative with values of at least a few $\mathrm{km\:s}^{-1}$. The most significant radial motion is found at radii between 400'' and 900'' (top right panel of Figure ~\ref{NGC2403}), and the strongest H\,{\sc i} mass flux of 3~M$_{\odot}\,\mathrm{yr}^{-1}$ is observed at 450'' (Figure~\ref{NGC2403}, bottom left panel). This mass flux  linearly declines towards larger radii.

\begin{figure}
 \begin{center}
	\includegraphics[width=1.04\widthbig]{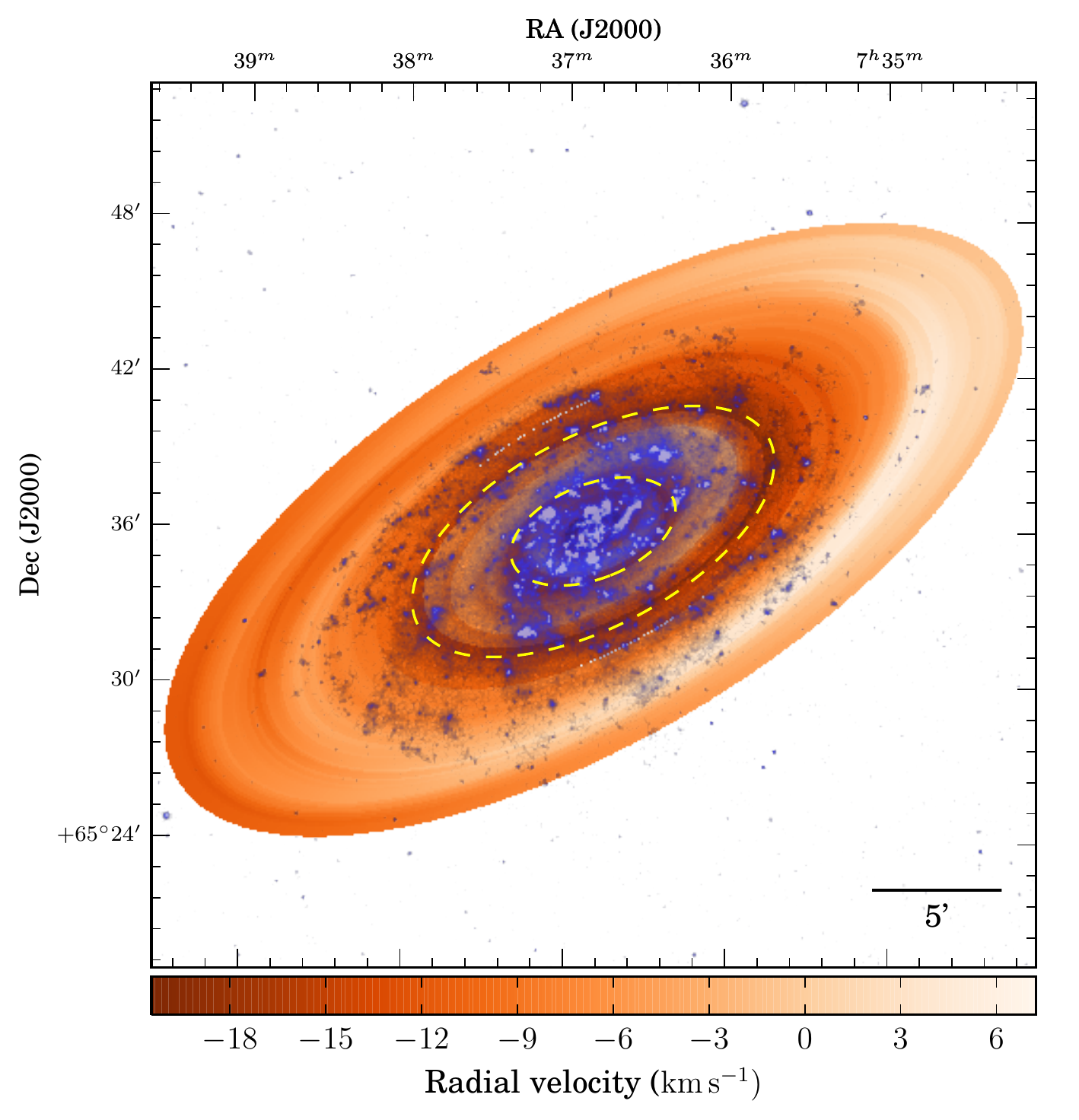}
 \end{center}
\caption{Combination of a reconstructed radial velocity field of NGC~2403 (orange shaded ellipses) and a GALEX FUV map tracing recent star formation. The yellow, dashed ellipses mark radii of 200'' and 460'', where the maximum inflow velocities are measured (compare to Figure~\ref{NGC2403}). It can be seen that most of the UV emission appears in the inner part of the disk, inside the region of peak inflow. This at least allows the interpretation that the gas moves inwards where it is transformed into stars. }
\label{NGC2403_Image}
\end{figure}

The quality of the fit we obtain for this galaxy is excellent. All parameters vary smoothly and adjacent rings show only very limited scatter. The residual deviation between data and our model (shown in Figure~\ref{NGC2403_Residuals}) are of the order of $\pm10\;\mathrm{km\:s}^{-1}$ and show no distinct pattern which would indicate a mismatch between fit and data. The flocculent spiral arms of NGC~2403 are just barely visible in the residual map. 
The terms $c_0$, $c_2$ and $s_2$, which indicate the deviation from a rotational symmetric velocity field, are fairly small inwards of 900'' (Figure~\ref{NGC2403}, bottom right part).  In the outskirts of the galaxy, there may be some lopsidedness. One can see some correlations between the different parameters around 200''. We cannot completely exclude that at this small radii a possible mismatch in e.g.\ the inclination induces artefacts in all parameters but these effects only appear in the inner part of the galaxy.

The star formation rate given in \cite{Leroy2008} is 0.38~M$_{\odot}\,\mathrm{yr}^{-1}$. This is of the same order as our minimum H\,{\sc i} inflow rate but nearly an order of magnitude lower than our maximum inflow rate of 3.5~M$_{\odot}\,\mathrm{yr}^{-1}$. 
If we follow our highly simplified model described in Section~\ref{continuity}, we would expect the radial mass inflow rate all radii to equal the star formation rate within that radius according to Equation~\ref{mass_conservation}. The bottom left panel of Figure~\ref{NGC2403} shows these two quantities and it is obvious that at most radii both quantities do not add up to zero. Also, we would expect the inflow to be monotonic with radius and to be highest at the outer edge of the H\,{\sc i} disk. This is clearly not the case. Not surprisingly, there are probably additional processes other than inflow and star formation influencing the H\,{\sc i} budget of NGC~2403.

One possible effect could be the ejection of ionised material from stellar winds perpendicular to the plane followed by recondensation and settling of that gas onto the plane at larger radii. This is often described as the galactic fountain model and could explain the additional mass flows. For NGC~2403 some indication for this was discovered by \citealt{Fraternali2002}. By taking position--velocity cuts parallel to the major and minor axis, they find substantial amounts of extraplanar H\,{\sc i} gas that rotates more slowly than the disk and is reported to show a net inward motion. The gas they detect lies roughly in the same radial range (200'' to 700'') as the main inflow found in our analysis. However, our method has only limited sensitivity to extraplanar gas because
its presence  will not substantially change the intensity weighted mean used to compute the moment maps. 
Another possible way to deviate from our stationary model is time-variable accretion, for instance if accretion happens via discrete events (e.g., large gas clouds).

The spatial correlation of inflow pattern and star formation is visualised in Figure~\ref{NGC2403_Image}. It shows a composite of the GALEX~FUV image and the reconstructed radial velocity field. 
The inflow has a maximum at 460" and decreases strongly at smaller radii. As described in Section~\ref{continuity}, this should lead to a pile-up of gas or enhanced star formation. The GALEX image shows clear clustering of star formation regions in this area. Most of the star formation takes place inside of 200", again a region where the magnitude of the 
inflow decreases. 
This spatial correlation may provide support, at least qualitatively, for a scenario in which H\,{\sc i} inflows are a key parameter regulating star formation in this galaxy. However, a quantitative comparison shows that the amplitudes of SFR and inflow do not match, which may not be surprising since a variety of other factors (time evolution, stellar winds, gas phase changes, etc.) likely play an important role as well.

Our observation for NGC~2403 clearly shows a qualitative correlation between star formation and radial H\,{\sc i} flows. This gives support to the picture that star formation may be regulated by inflows in NGC~2403. 
\citet{Meidt2013} analyse gravity-induced torques in the spiral galaxy M51, how they drive radial gas motions and how these are related to the local star formation efficiency. They find that radial motions seem to counteract star formation in this galaxy so that regions that experience higher torques show increased gas depletion times. This is the opposite sense of our finding here. M~51 and NGC~2403, however, have quite different properties and are not immediately comparable. The spiral arms in NGC~2403 are by far not as pronounced as in M~51 and the gravitational torques and dynamical driven inflows they might induce should be much lower and probably negligible. Our residual map (Figure~\ref{NGC2403_Residuals}) indeed shows no signs of spiral arm structure. It is therefore not necessarily expected that the ``dynamic pressure'' \citet{Meidt2013} describe suppresses star formation in NGC~2403 as it does in M~51. Unfortunately, due to its low inclination, we cannot analyse M~51 with the method presented in this study and directly compare the results.

\subsection{NGC 2841}
\label{SecNGC2841}

\begin{figure*}
 \begin{center}
	\includegraphics[width=\linewidth]{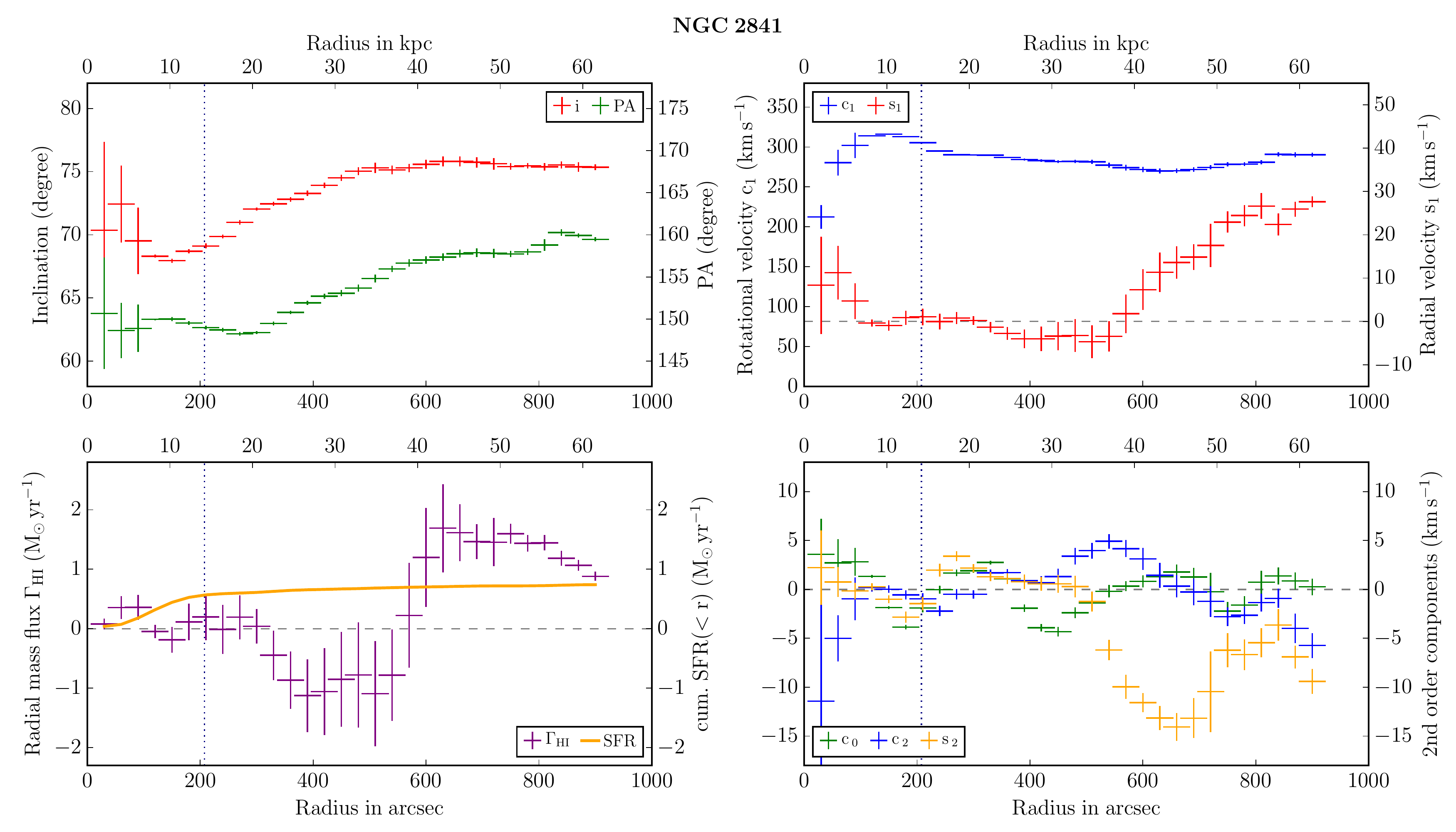}
 	\caption{
 	Disk geometry and Fourier parameters for NGC~2841. See caption of Figure~\ref{NGC2403} for technical details. 
 	The galaxy shows a fairly flat disk with a slight and smooth inclination increase in the inner half of the disk. 
 	Significant radial flows are only detected in the outer part, where gas seems to move away from the centre (red points in top right panel).}
 	\label{NGC2841}
 \end{center}
\end{figure*}

\begin{figure*}
 \begin{center}
	\includegraphics[width=.75\linewidth]{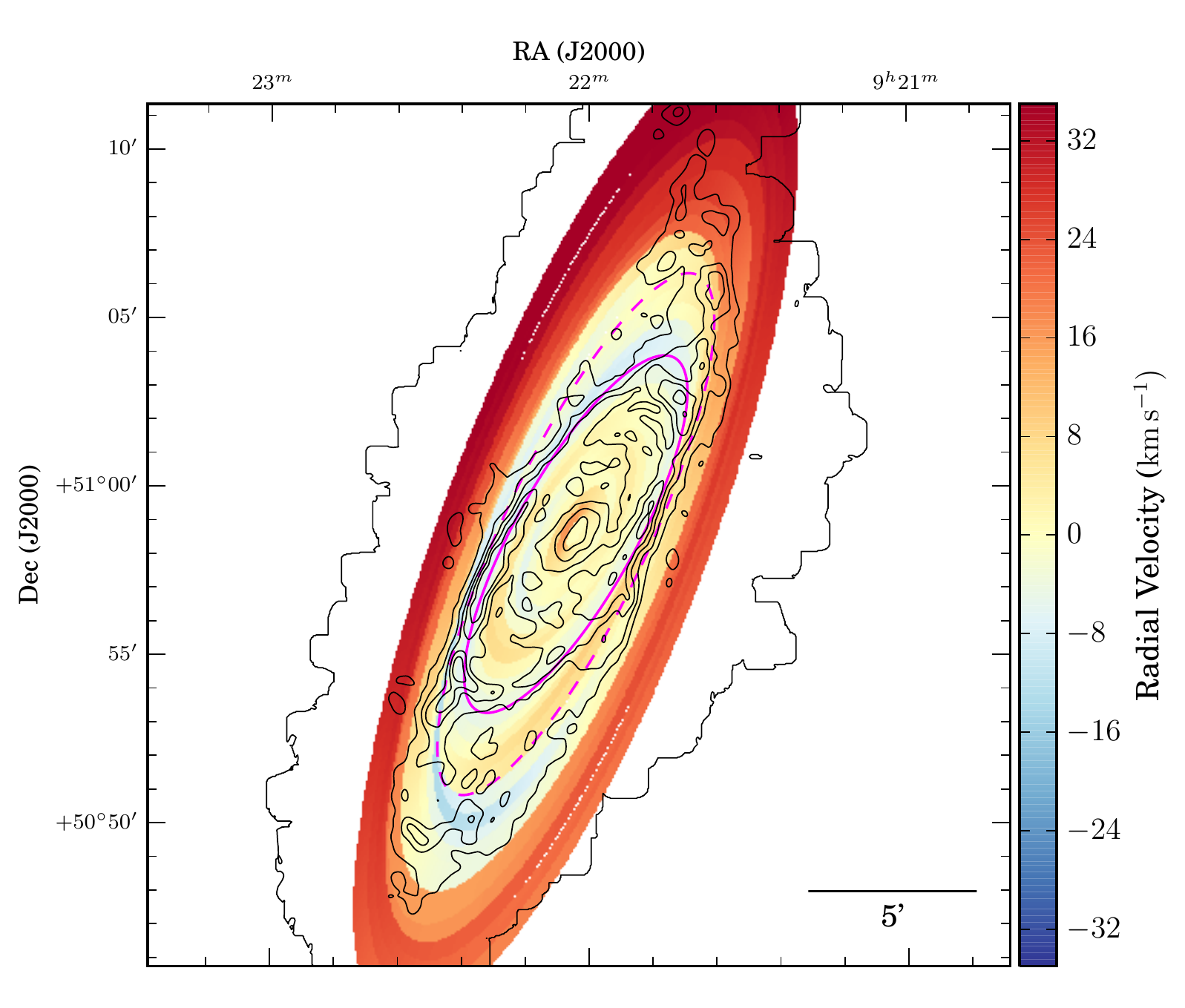}	
 	\caption{
 	Two-dimensional representation of our derived radial velocity field of NGC~2841.
 	The solid purple ellipse marks a radius of 360" at which the radial velocity has 
 	the strongest negative slope (compare to Figure~\ref{NGC2841}). At this radius we expect according to Equation~\ref{pile-up} an accumulation of gas.
 	The contours represent the H\,{\sc i} density distribution which indeed shows a pronounced ring structure with enhanced gas density exactly at this radius.
 	The dashed ellipse corresponds to 510" where we also find a fairly weak but in the southern half of the galaxy well visible negative slope in the radial velocity. This is as well spatially coincident with regions of increased H\,{\sc i} density.
 	It is therefore quite possible that radial gas motions and enhanced H\,{\sc i} surface density are physically related.
 	}
 	\label{NGC2841_Vrad+MOM0}
 \end{center}
\end{figure*}

NGC~2841 has a very extended H\,{\sc i} envelope, several times larger than the stellar disk and rotates counter-clockwise which means the north-eastern part of the disk is the near side.
Our fits to this galaxy are very stable and show a flat disk with 
very smoothly rising inclination and position angle (Figure~\ref{NGC2841}). 
The radial velocity (Fourier component $s_1$) is close to zero up to a radius of roughly 550''. We find a small inflow in the range from 350'' to 550'' which is consistent with the reported SFR of 0.74~M$_{\odot}\,\mathrm{yr}^{-1}$ \citep{Leroy2008}. 

From our SFR profile we find that most of the star formation is located within 200" while we detect inflow between 300" and 600".

Outside of 550'' the radial velocity rises to $25\;\mathrm{km\:s}^{-1}$ which corresponds to an H\,{\sc i} outflow of 2~M$_{\odot}\,\mathrm{yr}^{-1}$. The $s_2$ term indicates some lopsidedness of this radial flow.

From the appearance of very pronounced dust lanes in visual images, it is clear that the galaxy rotates counter-clockwise and the measured radial flow is indeed directed outward. From the morphological appearance one could speculate about tidal or merger effects, however NGC~2841 is not accompanied by any major galaxy and only surrounded by several dwarf galaxies (a NED\footnote{http://ned.ipac.caltech.edu} search yields only minor companions with $\mathrm{m_r>15~mag}$ within $3^{\circ}$ and $\pm400\,\mathrm{km\:s}^{-1}$).

It is interesting to compare the H\,{\sc i} distribution with the radial velocity field. A two dimensional representation of this is given in Figure~\ref{NGC2841_Vrad+MOM0}. It shows the radial velocity field we reconstruct from the determined Fourier parameters as described in Section~\ref{Fit_Procedure}. Overplotted are contours of the THINGS integrated intensity map. 

The H\,{\sc i} distribution shows a prominent ring at intermediate radii. If a radial mass flow causes some accumulation of gas we would expect this to happen as described in Section~\ref{continuity} at the location where the slope of $\mathrm{\Gamma_{rad}}$ is the most negative. From Figure~\ref{NGC2841} we see that this is the case at around 360". This radius is marked in Figure~\ref{NGC2841_Vrad+MOM0} with a purple ellipse which exactly coincides with the H\,{\sc i} ring. A second, but weaker feature like this can be seen at 510". 
This might indicate that our highly simplified model described in Section~\ref{continuity} even without any additional sink or source terms might actually provide a reasonable description  for the mass flow processes in NGC~2841.
Quantitatively, however, a mass flux with an amplitude of $\approx 1\:\mathrm{M}_{\odot}\,\mathrm{yr}^{-1}$ would have to act for a long time period to cause an enhancement in the H\,{\sc i} distribution of NGC~2841, which has a total H\,{\sc i} mass of $8.6 \times 10^9\:\mathrm{M}_{\odot}$.

\subsection{NGC 2903}
\label{NGC_2903}

\begin{figure*}
 \begin{center}
	\includegraphics[width=\linewidth]{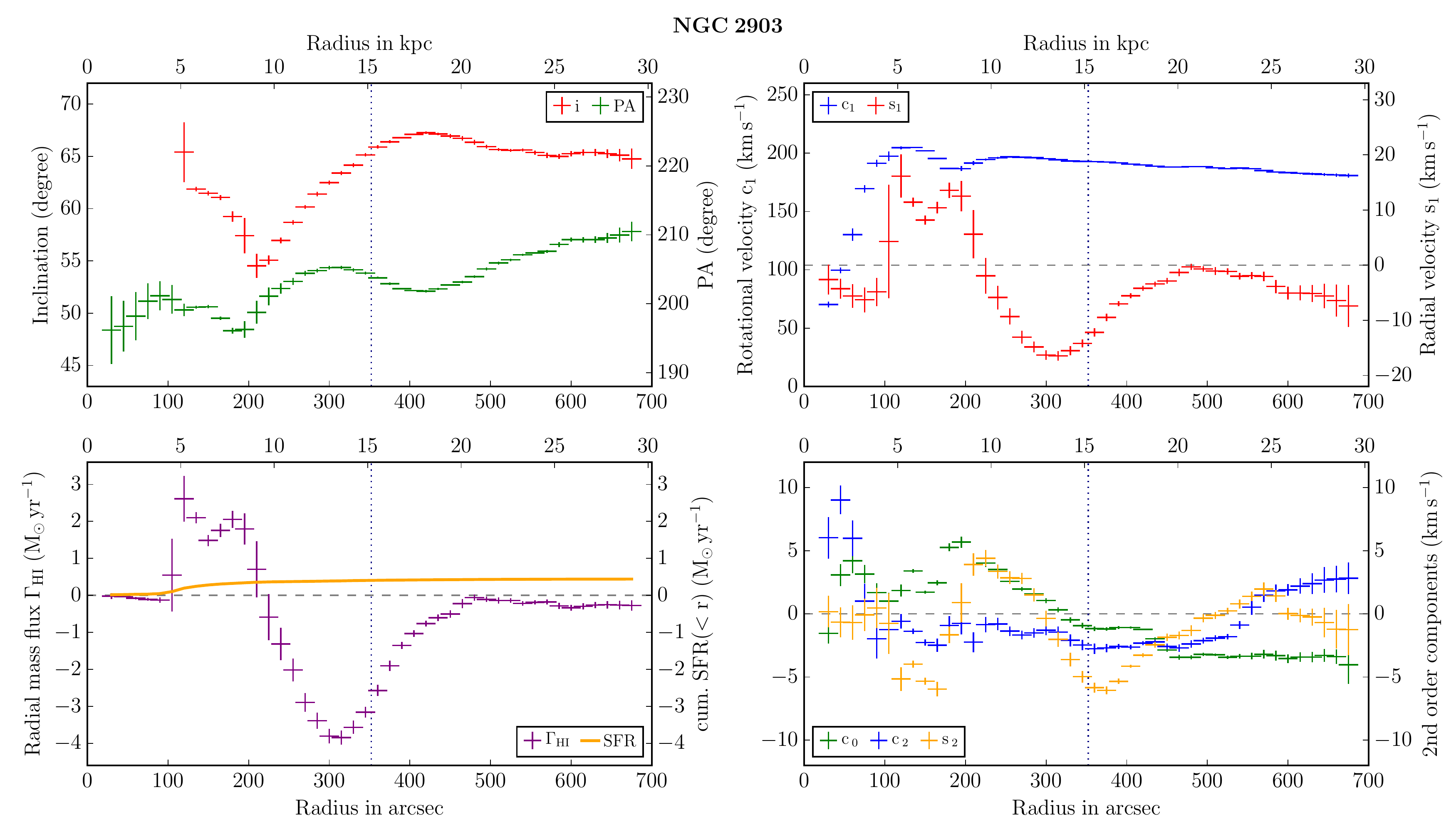}
	\caption{
	disk geometry and Fourier parameters of NGC~2903. The galaxy shows a strong bar in visual images which seems to complicate the determination of the disk geometry in the innermost part. It is also not clear if the dip in inclination around 200'' is a genuine feature or an artefact of the analysis (see discussion in Section~\ref{NGC_2903}). 
	The feature around 300'' seems to be related to internal streaming motions. 
	In the outskirts, the measured H\,{\sc i} mass inflow is only around 0.3~M$_{\odot}\,\mathrm{yr}^{-1}$. However, this is a still significant regarding the error estimates.
	}
	\label{NGC2903}
 \end{center}
\end{figure*}

NGC~2903 is rotating counter-clockwise and has complex kinematics. Visual images and the HERACLES CO-maps \citep{Leroy2009}  show a very clear bar structure which does not appear in the integrated H\,{\sc i} maps. 
The bar in conjunction with the solid body rotation in the inner regions prevents us from deducing proper fit results for radii smaller than 100'' and also causes a quite complicated structure outside of this radius (see Figure~\ref{NGC2903}).

At 200'', our analysis shows a sharp minimum of the inclination, followed by a steep increase from $55^{\circ}$ to $67^{\circ}$ within 200''.  This is close to the fastest inclination change rate we expect from Equation~\ref{i_crit} to still yield reliable fit results. We take this as an indication that the fit in this region is likely to  be not very reliable. This dip in inclination furthermore coincides with two regions of very low H\,{\sc i} flux, basically holes in the H\,{\sc i}~disk. In general, due to the complex kinematic structure of this galaxy the fit results for radii below 250'' could be affected by artefacts and should probably not taken at face value.

The largest feature in our $s_1$ curve is a massive inflow signature around 300''. 
Similar to NGC~2841, we can assess whether the radial gas motions could lead to the accumulation of gas in the ring structure visible in the integrated intensity map. Based on Equation~\ref{pile-up}, we would expect a pile up of gas at a radius between 200" and 300" due to the very steep negative slope of $\Gamma_\mathrm{rad}(R)$ but this region shows a rather low H\,{\sc i} surface density. The ring structure is located at a radius of $330"\pm20"$. This does not correspond to any particular feature in the radial flux profile. The peak inflow at 300" also does not coincide, but is just inside the ring structure.
In contrast to NGC~2841, our simple model described in Section~\ref{continuity} is obviously not able to describe the relation between radial motions and H\,{\sc i} density in NGC~2903. 
We speculate that the ring structure might instead be related to more complex internal streaming motions, probably associated with the galactic bar, the effects of which are likely not captured correctly by our 2nd second-order analysis \citep{deBlok2008}.
No further galaxy in our sample shows gas density features in sufficient detail to allow a straight forward comparison of radial motions to H\,{\sc i} density. The absence of gas accumulation can always be explained by the low amplitude of order $1\:\mathrm{M}_{\odot}\,\mathrm{yr}^{-1}$ of the radial motions.

Further out beyond 500", the kinematic structure of the disk seems less complicated and the velocity measurements indicate a quite substantial inward motion. However, due to the low gas surface densities at these radii, the corresponding H\,{\sc i} mass flow rate is only 0.3~M$_{\odot}\,\mathrm{yr}^{-1}$. This is quite small but similar to the SFR reported by \citet{Popping2010}, who report values ranging from 0.4~M$_{\odot}\,\mathrm{yr}^{-1}$ derived from UV measurements up to 1~M$_{\odot}\,\mathrm{yr}^{-1}$ from H$\alpha$ observations. Our own SFR measurement is consistent with \citet{Popping2010} but also shows that most of the star formation happens inside of 150". Since our kinematic analysis does not yield reliable results in this region, a direct comparison is not possible. However, the data are still compatible with a scenario in which star formation inside of 150" is fueled by the inflow visible in the disk beyond 500".

\subsection{NGC 3198}

\begin{figure*}
 \begin{center}
	\includegraphics[width=\linewidth]{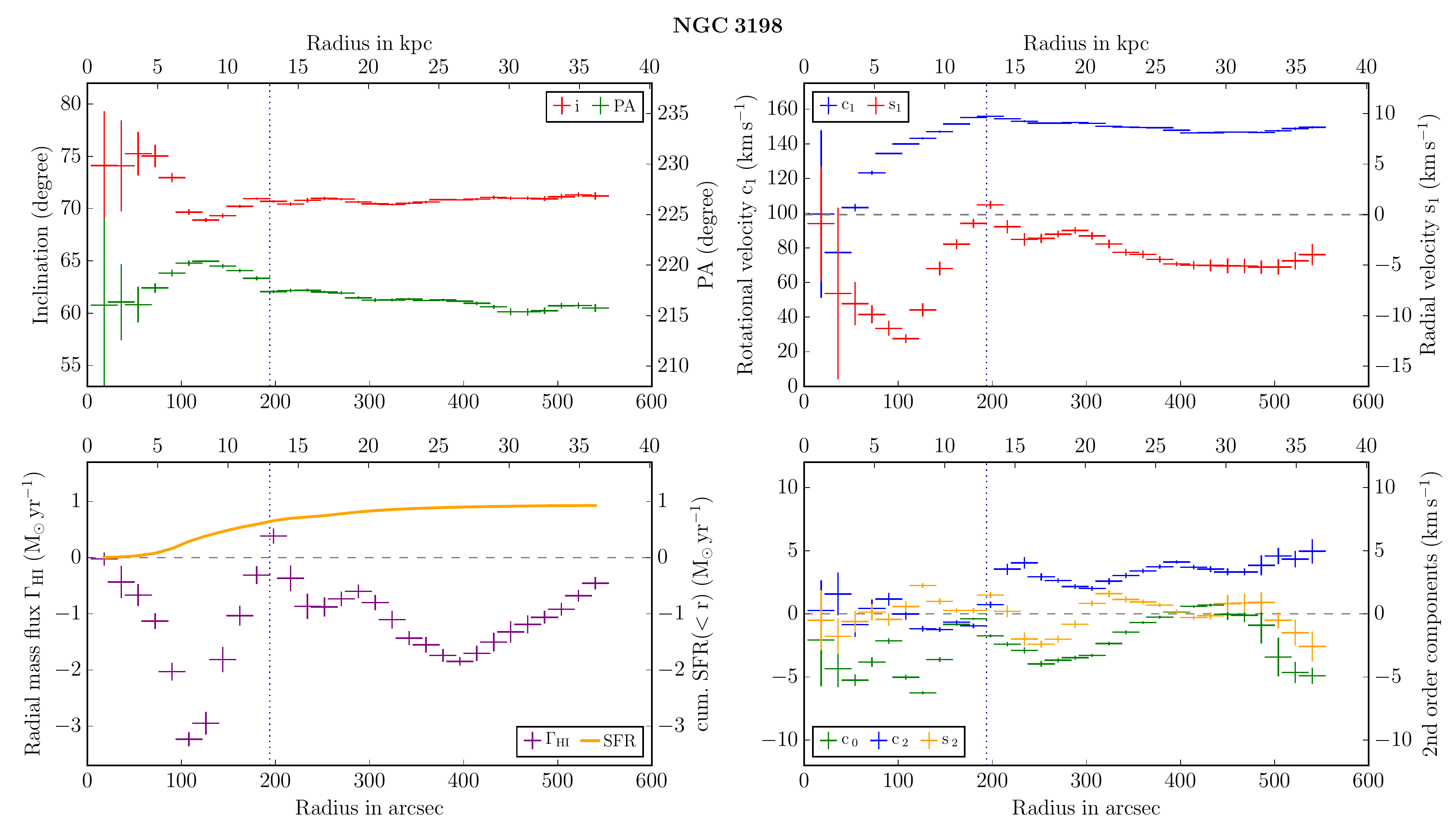}
	\caption{The results for NGC~3198 show a flat and smooth disk with some inflow of nearly constant velocity in the outer half of the galaxy.}
	\label{NGC3198}
  \end{center}
\end{figure*}

NGC~3198 also rotates counter-clockwise and the south-east part is the nearside. The galaxy has a flat disk without any substantial changes in inclination or position angle outside of 150'' (Figure~\ref{NGC3198}). 

The radial velocity we measure is always negative outside of 200" and we find significant inflow between 0.7 and 2~M$_{\odot}\,\mathrm{yr}^{-1}$. The average inflow outside $r_{25}$ is 1.1\,M$_{\odot}\,\mathrm{yr}^{-1}$. This agrees quite well with the star formation rate of 0.93~M$_{\odot}\,\mathrm{yr}^{-1}$ reported by \citet{Leroy2008}.
Also the radial distribution is roughly consistent. Most of the star formation happens  between 5~and~15~kpc. In this region we find a substantial modulation of the inflow, probably caused by the star formation process, but outside of this radius we have a roughly constant inflow with an amplitude similar to the star formation rate. Therefore the overall trend fits quite well in the picture that star formation is fueled by inflows for this galaxy.

If we compare our results for the $s_1$ component to the values obtained by \cite{Schoenmakers1997} and \cite{Trachternach2008}, we find them to be very similar. We can reproduce their results by running a two-step analysis. Even if no radial flow is taken into account in the tilted ring fit, the subsequent Fourier decomposition shows clear signs of radial inflow with at least 50\% of the amplitude found in the simultaneous fit. This is somehow surprising since for NGC~2403 neither \cite{Schoenmakers1997} or \cite{Trachternach2008} or our two-step fit could recover substantial radial inflows. For a detailed discussion of different fitting schemes and validations on mock velocity fields compared to Section~\ref{TwoStepFit}. Clearly some property in the NGC~3198 velocity field is different compare to NGC~2403. As we show in Figure~\ref{NGC3198}, the $s_1$ term is relatively small and the terms $c_0$, $c_2$ and $s_2$ have comparable amplitude. This means, the asymmetry of the velocity field is of comparable amplitude as the overall net inflow. It may be possible that the inflow in this galaxy occurs in such a way that the tilted ring fit cannot adjust $i$ and $PA$ to fully compensate for the inflow. Indeed, inspection of the H\,{\sc i} intensity map (available from the THINGS web page) shows an overall asymmetry with an extended north-east arm.

\subsection{NGC 3521}

\begin{figure*}
 \begin{center}
	\includegraphics[width=\linewidth]{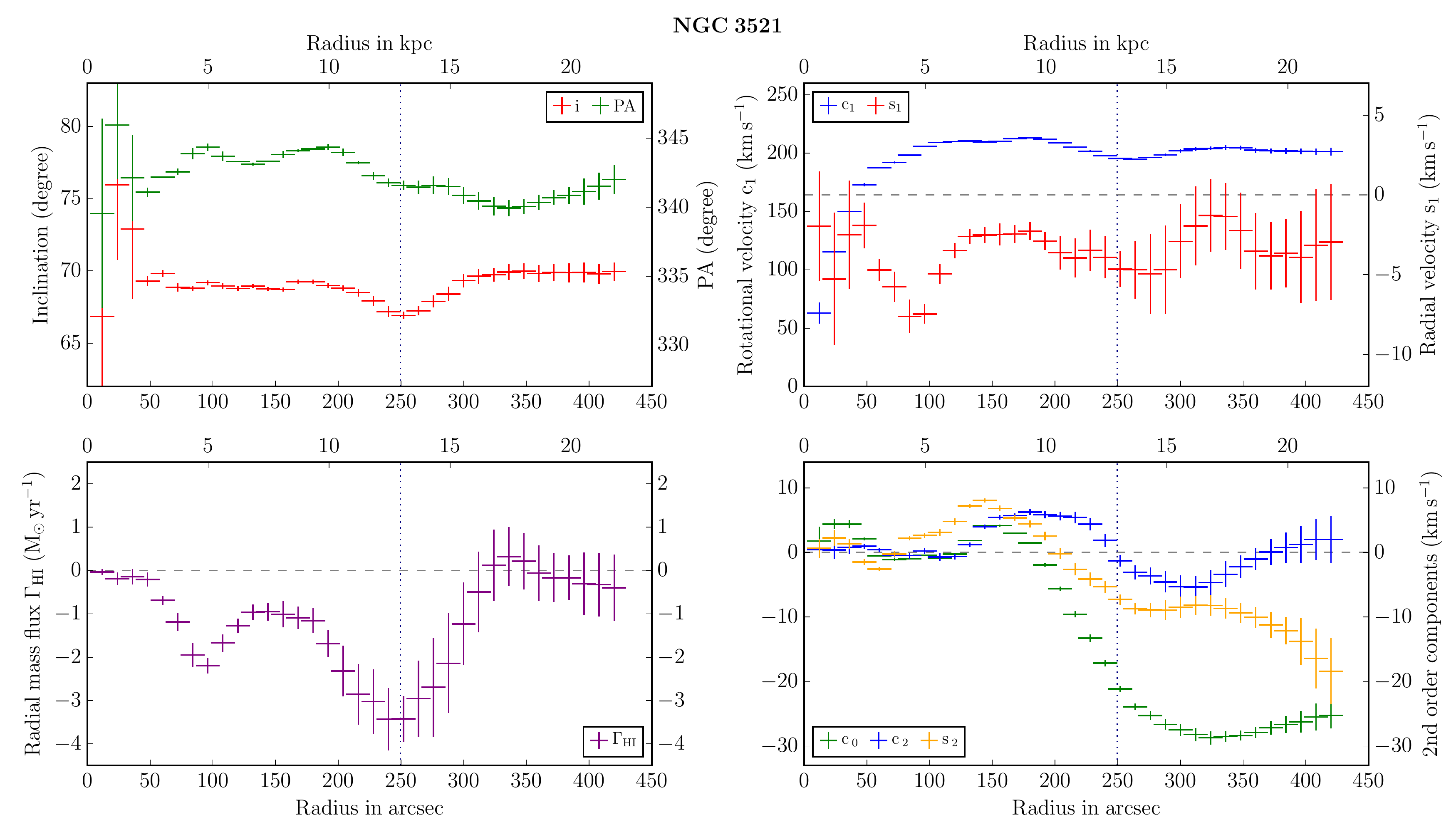}
	\caption{The disk geometry of NGC~3521 seems quite flat with some radial inflow mostly at intermediate radii. An odd feature is the $c_0$ component which drops to a value of $-30\;\mathrm{km\:s}^{-1}$ outside of 200''. This indicates an extreme lopsidedness or hints at other peculiarities in the velocity field. For NGC~3521 no GALEX~FUV image was available to derive a SFR profile.}
	\label{NGC3521}
 \end{center}
\end{figure*}

NGC~3521 is rotating counter-clockwise and its western part is closer to us. Inclination, position angle and rotation velocity are fairly constant. The radial velocity seems to be of low amplitude, as depicted in Figure~\ref{NGC3521}. The star formation rate reported by \citealt{Leroy2008} is 2.1~M$_{\odot}\,\mathrm{yr}^{-1}$. No GALEX FUV image was available in the data archive to derive a SFR profile (for the sake of consistency and comparability we refrain from constructing this profile based on other SFR tracers). 

We find a rather unusual behaviour for the $c_0$ term, which starts to drop at about 200'' and levels off at a value of $-30\;\mathrm{km\:s}^{-1}$. The term should represent a strong lopsidedness of the radial velocity or a significant change of the systemic velocity with radius. Since the latter seems unreasonable, we conclude that this term represents lopsidedness. This in consequence would imply a strong east-west gradient in the radial velocity field. Gas then would enter the galaxy on the western side with roughly $30\;\mathrm{km\:s}^{-1}$, cross the disk and leave again on the eastern side, which altogether seem quite an unlikely scenario too.

However, the overall galaxy appearance is indeed quite asymmetric. Images at optical wavelength show numerous well pronounced dust lanes and a very extended north-west spiral arm. This arm is also visible in the 21~cm line as extension of the disk. In addition, also a south-east lobe exists which  is even more extended. It shows very high fit residuals of up to $+60\;\mathrm{km\:s}^{-1}$. This relatively localised feature introduces the peculiar $c_0$ values to our fit. We know that NGC~3521 underwent extensive merger activity in the past. \citet{Martinez-Delgado_2010} present deep surface photometry which shows several stellar streams and an extended envelope of stars around the galaxy. Also, \citealt{deBlok2008} found that a substantial amount of gas is not moving within the plane of the galaxy. This clearly affects our fit.

Given the complex dynamic structure of NGC~3521 and its recent merger history, the results of our fitting procedure need to be considered with caution. The derived  fluxes need not necessarily be interpreted as large-scale inward mass motion. Deciphering the kinematics of such a disturbed system most likely is beyond the capabilities of the method, as it relies on the assumption of a relatively unperturbed disk with only mildly varying geometry.

\subsection{NGC 3621}

\begin{figure*}
 \begin{center}
	\includegraphics[width=\linewidth]{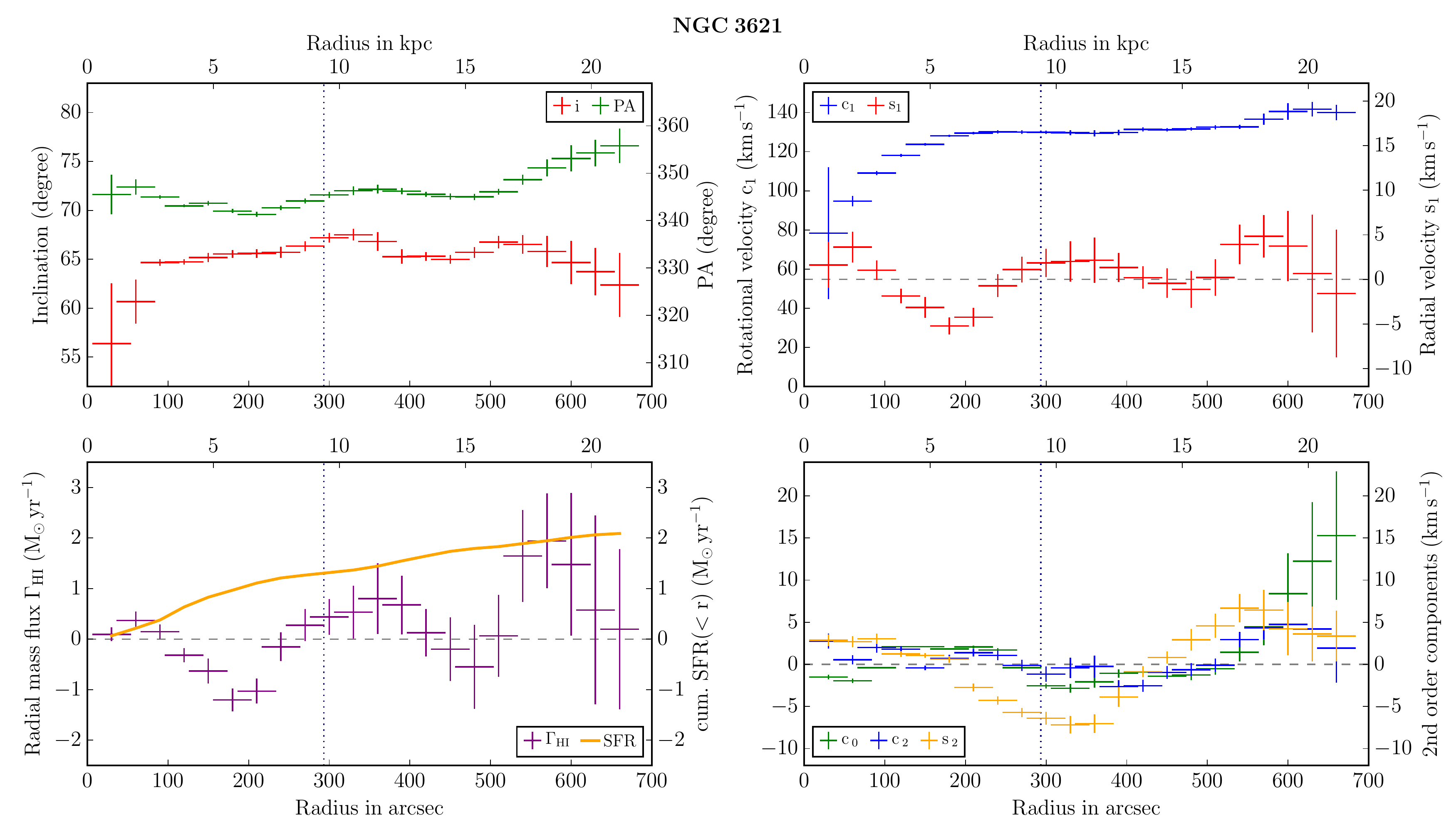}
	\caption{Fit results for NGC~3621. The disk is flat, as is the rotation curve. The radial velocity wiggles around zero without clear signs of significant radial motions.}
	\label{NGC3621}
  \end{center}
\end{figure*}

The analysis for NGC~3621 reveals a relatively simple disk with fairly constant inclination and position angle. This is somewhat surprising, because the galaxy looking clumpy and inhomogeneous in the integrated intensity map. The visual impression could be influenced by the fact, that 
an extra gas component in the form of a stream or a warp in the line of sight crosses the main disk
and creates some foreground confusion \citep{deBlok2008}. The galaxy rotates clockwise and the rotation curve is completely flat outside of 200'' (Figure~\ref{NGC3621}). The measured radial velocity is of low amplitude  and varies in the range of $\pm5\;\mathrm{km\:s}^{-1}$. 
\citealt{walter2008} report a high SFR of 2~M$_{\odot}\,\mathrm{yr}^{-1}$. Our coarse SFR estimate from GALEX FUV images is substantially lower (0.5~M$_{\odot}\,\mathrm{yr}^{-1}$) but as mentioned, our aim is to reproduce the radial distribution not the galaxy-integrated amplitude of the star formation activity. Figure~\ref{NGC3621} therefore shows the profile we measure scaled to the total SFR reported by \citet{walter2008}.  Some slight correlations between the radial H\,{\sc i}  flow and the SFR profile might be visible, in particular at 100" where a substantial amount of the star formation takes place and the low-level inflow we measure decreases. However, we do not find signs of large-scale inflow and the recent star formation in NGC~3621 seems to consume the existing gas inventory of the galaxy and is at the moment not fueled from outside in substantial amounts.

\subsection{NGC 5055}

\begin{figure*}
 \begin{center}
	\includegraphics[width=\linewidth]{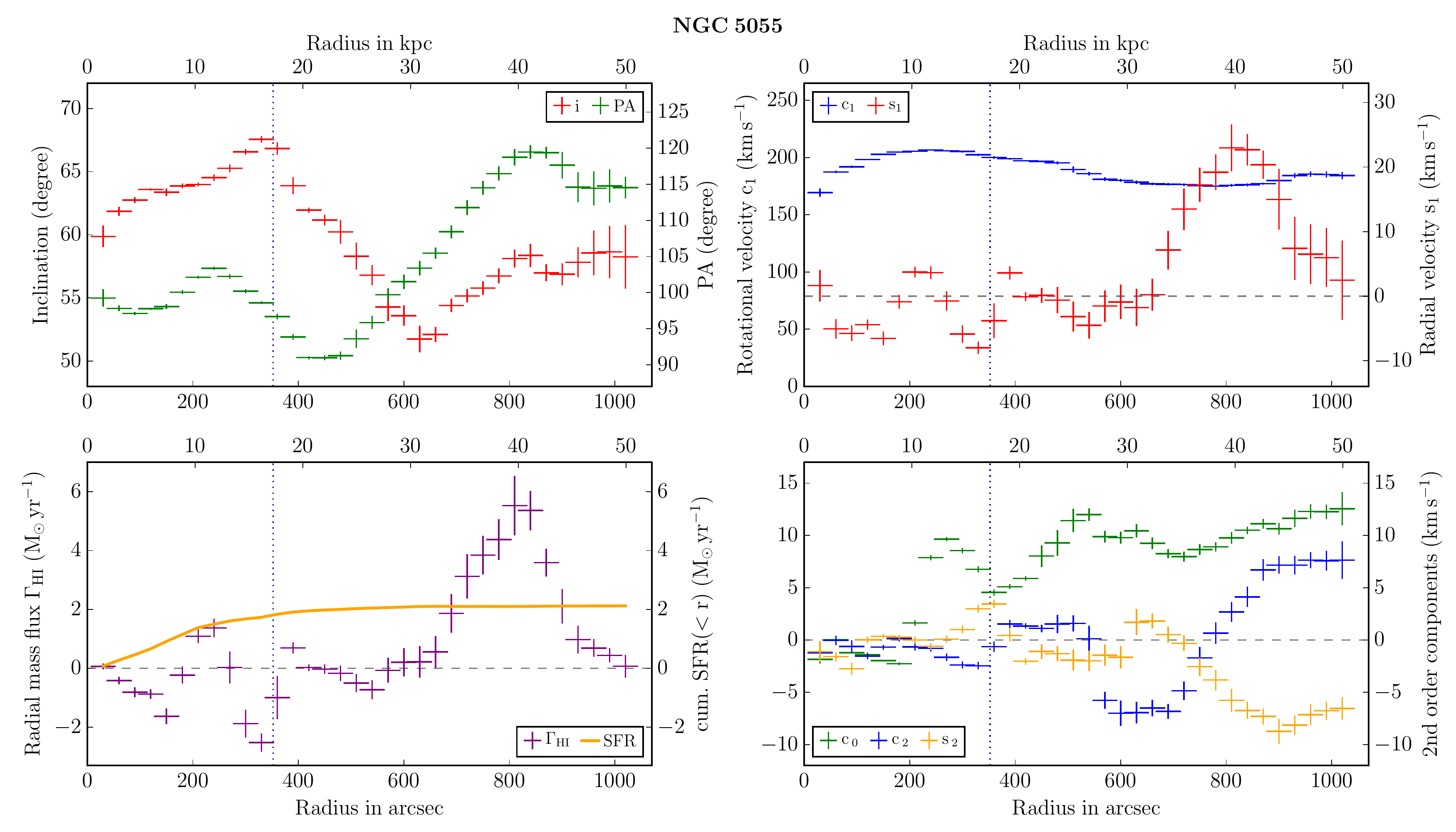}
	\caption{disk geometry and Fourier parameters of NGC~5055. The geometry of the disk shows significant structure compared to the other galaxies in our sample. None of the analysed galaxies shows as much  change in inclination and position angle as NGC~5055. We speculate that this is due to an interaction with a dwarf galaxy. A corresponding stellar stream reported by \citet{Chonis2011} intersects the disk at a radius of roughly 600''. This coincides with the radius at which we see a distinct minimum of the inclination and a strong change of position angle. Inside of 600'' we find no significant radial motions, further out we  find gas moving outward, peaking at a radius of 800'' and H\,{\sc i} mass flow rates of up to 6~M$_{\odot}\,\mathrm{yr}^{-1}$.
	}
	\label{NGC5055}
  \end{center}
\end{figure*}

NGC~5055 exhibits a very complex disk geometry, as indicated in Figure~\ref{NGC5055}. The southern part of the disk is closer to Earth and rotation occurs in a clockwise direction. In the integrated H\,{\sc i} map as shown by \citet{walter2008} one can clearly distinguish between an inner star-forming part of the galaxy (SFR given by \citealt{Leroy2008} 2.1~M$_{\odot}\,\mathrm{yr}^{-1}$, for the radial distribution, compare to Figure~\ref{NGC5055}) and an outer diffuse ring with a noticeably different inclination angle. The kinematic analysis yields an inclination which first rises to above $65^{\circ}$ at 380'', then declines to $50^{\circ}$ at 650'' and then rises again to slightly below $60^{\circ}$. The position angle also changes. The most significant feature is a steep rise by $30^{\circ}$ between 500'' and 800''. For the radial velocity we find values around zero up to a radius of 600''. Outside of this, in the region of the H\,{\sc i} ring, it rises to $20\;\mathrm{km\:s}^{-1}$ and declines again towards the outermost part of the galaxy. This corresponds to an outward peak mass flow of 6~M$_{\odot}\,\mathrm{yr}^{-1}$. Inside of 600'', no significant mass flow is detected. The asymmetry terms show quite high and varying values.

Even if the rise of inclination outside of 600'' is quite steep, it is only about 50\% of the critical change rate given in Equation~\ref{i_crit}. Therefore, the measured $s_1$ term should be robust and we consider the H\,{\sc i} mass flow to be real. From dust lanes we can infer that the galaxy rotates clockwise and the radial flow is directed outward. \citet{Chonis2011} report the detection of a stellar stream in the outskirts of NGC~5055 via deep optical surface photometry. The stream is inclined with respect to the disk, but not associated with any H\,{\sc i} structure. NGC~5055 is part of the M51-group and accompanied by dwarf galaxies are known \citep{Battaglia2006}. The detection of the stream is a clear evidence for a recent merger event. From the geometry of the stream presented by \citet{Chonis2011} one could estimate that it passed through the disk of NGC~5055 at a radius of roughly 600''. This matches strikingly well with the minimum of the inclination, and it is also well within the regime of the steep change in position angle. It seems that the merger has perturbed the disk of NGC~5055 and left an imprint in the H\,{\sc i} kinematics. Even if the exact dynamics are unknown, it is very likely that the complex disk geometry we find is the direct result of this recent merger event. 

There is also a dark dust filament visible in front of the southern half of the stellar disk. It has a quite luminous H\,{\sc i} counterpart which is easy to identify in the integrated intensity map. From the visual impression of the velocity map, the filament seems to move out of the galactic plane and at a different orbital velocity than the rest of the disk. This probably causes the bump in the inclination at around 350'' since the fit cannot differentiate between the disk velocity and the motion of the filament and fits an average along the line-of-sight.

\subsection{NGC 6946}

\begin{figure*}
 \begin{center}
	\includegraphics[width=\linewidth]{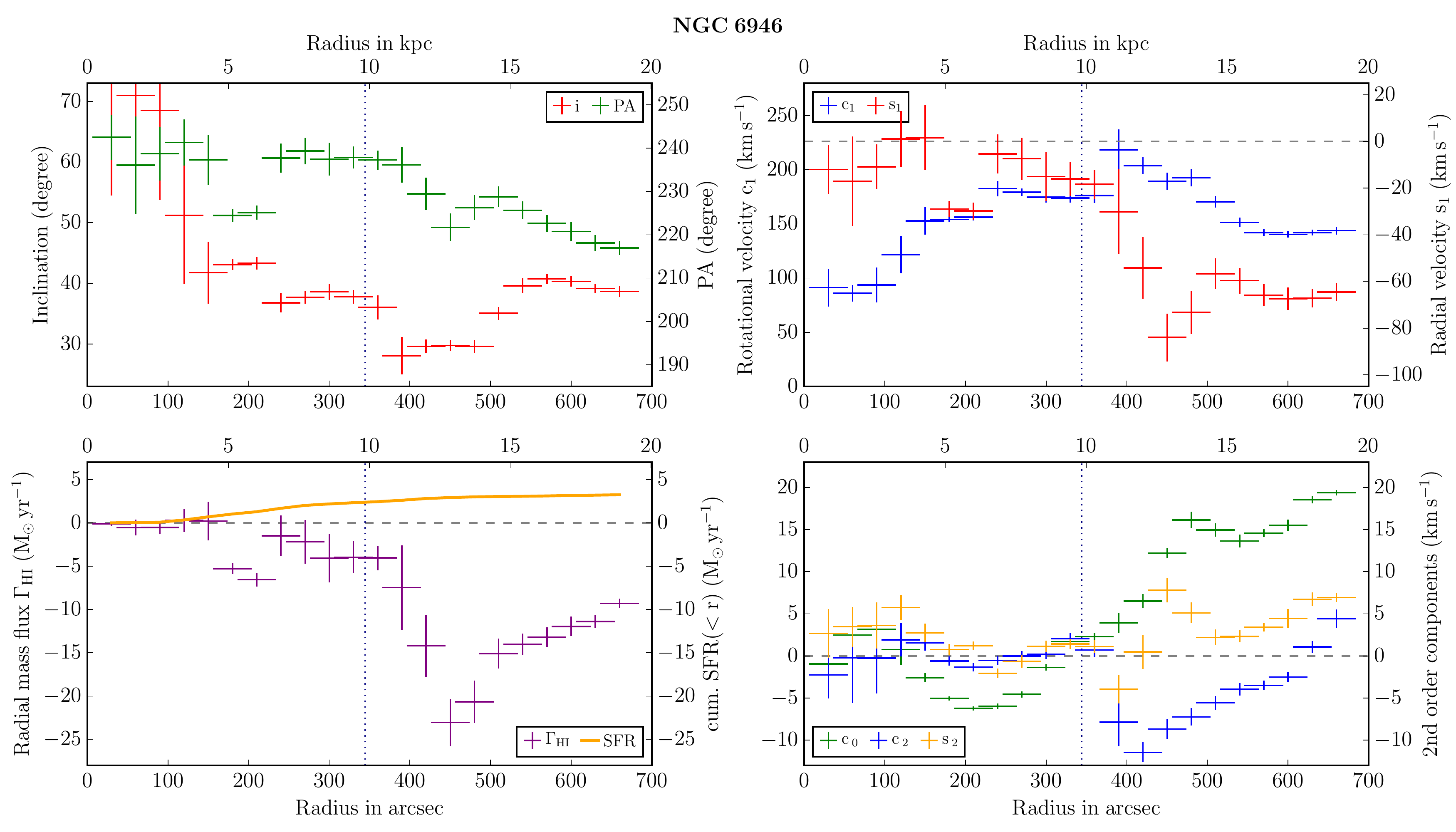}
 	\caption{
	Results of the analysis for NGC~6946. The galaxy has a low inclination of about $35^{\circ}$ which makes a kinematic analysis difficult and leads to large uncertainties. The determined radial velocity of $\approx60\;\mathrm{km\:s}^{-1}$ 
	translates into an H\,{\sc i} inflow of 15~M$_{\odot}\,\mathrm{yr}^{-1}$. While this represents an extreme mass flow, NGC~6946 has a high SFR of 3.2~M$_{\odot}\,\mathrm{yr}^{-1}$ and a high and extended molecular gas content. 
	}
	\label{NGC6946}
 \end{center}
\end{figure*}

For NGC~6946 we derive a fairly low inclination of $\sim 35^{\circ}$ with substantial scatter from ring to ring. Similar results are reported by \citet{deBlok2008}. The low and varying inclination makes the fit quite uncertain since the majority of parameters are scaled by $\sin(i)$.
The fit only converges if we choose quite wide rings (increase in 30'' steps) and we consider it quite likely that the actual uncertainties are larger than the formal error bars shown in Figure~\ref{NGC6946}.

We determine a moderately varying $PA$ and clockwise rotation and find the nearside of the disk to be the north-west part. The inward radial velocities are very high and in excess of $-60\;\mathrm{km\:s}^{-1}$, which is roughly one third of the rotational velocity. Also, the radial velocity field shows a quite substantial lopsidedness with a difference between north-west and south-east edge of around $30\;\mathrm{km\:s}^{-1}$. Even if we do not take the two rings at 450'' and 480'' into account, NGC~6946 shows an extremely high H\,{\sc i} mass inflow of 10 to 15~M$_{\odot}\,\mathrm{yr}^{-1}$ in the range from 400'' to 650''. Our fit results are stable and the residuals show no obvious feature that would indicate problems with the fit beyond the large statistical errors. Nevertheless, we express some concern that the derived magnitude of the inflow could be larger than what is physically reasonable and may be exaggerated due to the low inclination of the galaxy.

However, NGC~6946 does exhibit exceptional properties. The star formation rate is relatively large with with a value of  3.3~M$_{\odot}\,\mathrm{yr}^{-1}$ \citep{Leroy2008}. The gas-to-stellar mass ratio is also high. For the atomic gas, a value of 20\% is reported. For the molecular gas, the number is 12\%, which is the largest value  in the entire THINGS sample. 
From the GALEX FUV images we find that a substantial part of the
total star formation takes place inside of $\mathrm{r_{25}}$, well inside
the region where we detect the large inflows. The amplitudes of inflow and
SFR roughly match, however the uncertainty in the inflow measurements
prohibits a detailed comparison.

The general picture we get for NGC~6946 fits the simulations  by \citet{Marinacci2013} very well.
They ran eight cosmological hydrodynamical zoom simulations of Milky Way-sized halos using the moving mesh code AREPO to study the evolution of disk galaxies. 

They usually find quite fast inflow with radial velocities nearly matching the circular velocity in the outer part of the disk. The gas in the simulated disks is supported by rotation only up to the edge of the luminous disk and outside of this it is highly sub-keplerian. Although these high radial velocities fit qualitatively to our findings for NGC~6946, their simulations generally predict too small H\,{\sc i} disks compared to other spiral galaxies. The authors explain that their objects include far less gas than would be realistic and we assume that this could affect the size and dynamics of the simulated gas disks.

If the magnitude of the derived inflow is indeed correct, the inflow rates would be uniquely high in our sample. However, given the low inclination of NGC~6946 and the large ring-to-ring variations in the fit, we may expect these inflow rates to be upper limits and presumably should not be taken at face value.

\subsection{NGC 7331}

\begin{figure*}
 \begin{center}
	\includegraphics[width=\linewidth]{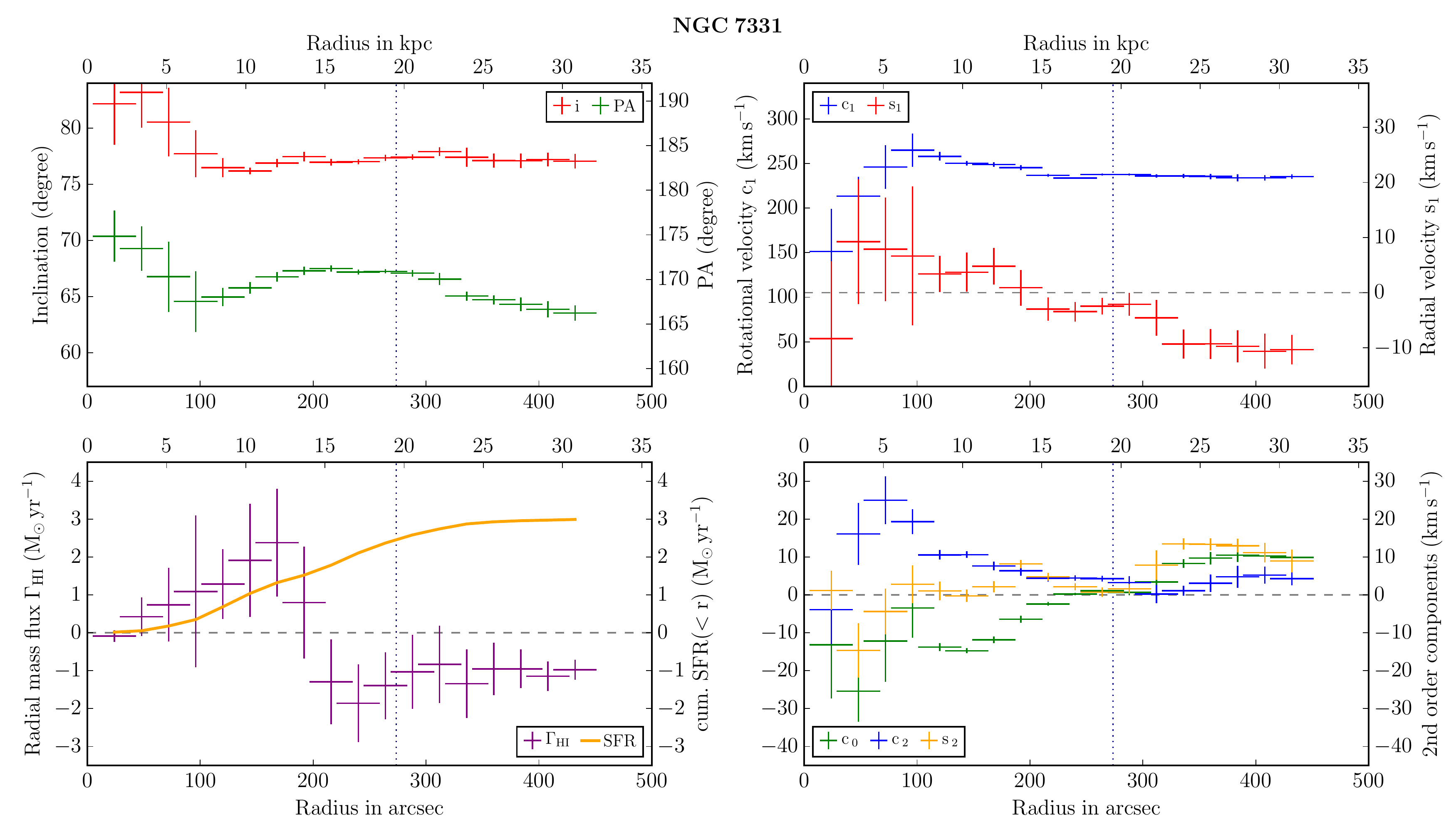}
	\caption{disk geometry and Fourier parameter for NGC~7331. The disk is flat and outside of 300'' we find $10\;\mathrm{km\:s}^{-1}$ inflow. Due to the relatively low H\,{\sc i} density this translates into an H\,{\sc i} mass flux of only 1~M$_{\odot}\,\mathrm{yr}^{-1}$.
	}
	\label{NGC7331}
 \end{center}
\end{figure*}

NGC~7331 is rotating clockwise and the western part of the disk is closer to Earth. Our analysis yields a fairly constant inclination and position angle as well as rotation velocity (Figure~\ref{NGC7331}).  We find some inflow of H\,{\sc i} at the level of 1~M$_{\odot}\,\mathrm{yr}^{-1}$ outside of $r_{25}$ and outflow inside of it. The radial velocity is close to zero in the inner part and negative at around $-10\;\mathrm{km\:s}^{-1}$ in the outer part of the  disk. The asymmetry terms are of similar magnitude. In the reconstructed radial velocity field one sees that inflow predominantly happens in the northern part of the galaxy. In the southern half, the radial velocity does not substantially deviate from zero. The integrated H\,{\sc i} map shows a fairly extended lobe on the northern end of the galaxy and a small on in the south. One possible explanation could be that a relatively localised gas stream flows inward from the circumgalactic medium and reaches the  northern part of the galaxy. Unfortunately, the high inclination of NGC~7331 makes it difficult to study the structure of the northern lobe in more detail, and the observations are not sensitive enough to trace the H\,{\sc i} gas further out.

The Leroy SFR for NGC~7331 is nearly as high as for NGC~6946, but the measured H\,{\sc i} inflow rate of about 1~M$_{\odot}\,\mathrm{yr}^{-1}$ is more than an order of magnitude lower. However, all star formation takes place inside of 25~kpc (compare to Figure~\ref{NGC7331}) and we find basically constant inflow outside of 15~kpc. Despite the mismatch in amplitude this gives a roughly consistent picture of mass transport and star formation.

\subsection{NGC 925}

\begin{figure*}
 \begin{center}
	\includegraphics[width=\linewidth]{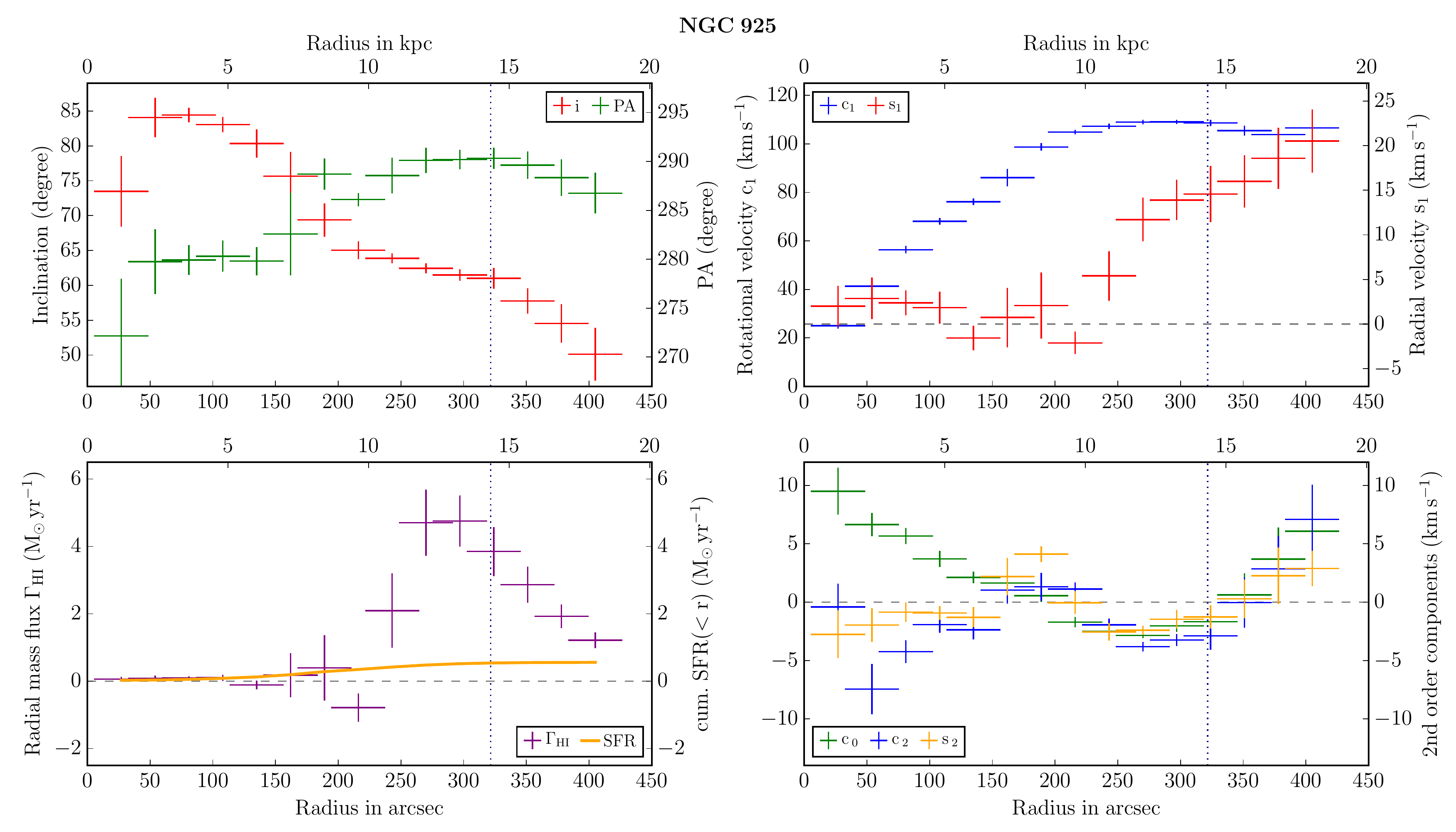}
	\caption{
	The results for NGC~925 indicate a linearly rising rotation curve which complicates the proper determination of the inclination in the inner half of the H\,{\sc i} disk (see discussion in Section~\ref{solid_body}). 
	Outside of 200'' our analysis indicates strong outflows.
	}
	\label{NGC0925}
 \end{center}
\end{figure*}

For NGC~925, we find that the galaxy is most likely rotating clockwise and that the northern part of the disk is closest to us. There is a nearly linear drop of inclination from above $85^{\circ}$ in the centre to $50^{\circ}$ in the outskirts. At the same time the rotation velocity shows a nearly linear rise out to 250'' and then saturates at $\sim 110\;$km$\,$s$^{-1}$ (Figure~\ref{NGC0925}). Optical images clearly show a bar in the galaxy centre. As described in Section~\ref{solid_body}, solid body rotation makes it impossible to disentangle inclination and rotation, and therefore the inferred disk geometry inside of 250'' may be incorrect. Outside of this radius, however, our fit results should be robust.

The amplitude of the radial velocity is quite high. Outside of 300'' it rises to a maximum of $20\;\mathrm{km\:s}^{-1}$. This results in an H\,{\sc i} outflow of nearly 5~M$_{\odot}\,\mathrm{yr}^{-1}$ at around 300''. The integrated H\,{\sc i} map clearly shows two extended spiral arms, probably related to interactions or accretion events. 
\citet{Heald2011} report the detection of an extended H\,{\sc i} envelope with complicated streams caused by a minor merger in the HALOGAS survey. They even find a very faint optical counterpart which could be the remnant of a gas-rich dwarf galaxy that was accreted by NGC~925. 
The fact that we derive outward instead of inward motion is surprising, but the overall picture slightly resembles that of NGC~2841 (Section~\ref{SecNGC2841}).

\section{Conclusions}
\label{sec:conclusions}

In this kinematic study we analysed the H\,{\sc i} velocity field of ten disk galaxies from the THINGS survey and 
investigated the presence of radial inflows, which might supply the star forming regions in the inner parts with fresh gas and thereby allow continued star formation. 

Given that the gas depletion timescale in most spiral galaxies is short and of the order of $\sim 2$~Gyr it is a long standing problem of galaxy evolution to explain how star formation activity is sustained over cosmological timescales. If we discard the possibility that most galaxies are observed on the verge of running out of gas, then they must be continuously fed by fresh material. Our aim is to search for signatures of this process. 

We started our investigation of the dynamics of H\,{\sc i} gas in nearby disk galaxies with the analysis of existing tilted ring fitting methods and the assessment of their statistical reliability. We found that most two-step fitting schemes are unable to recover the full radial inflow in the velocity field and as a consequence developed a new Fourier decomposition scheme that includes second order terms 
of rotational and radial velocity 
and simultaneously determines the position angle and inclination angle as functions of radius. By applying it to mock velocity data we demonstrated that the new method is more stable and less susceptible to systematic errors. It is well suited to reliably detect radial flows of material through the disk for galaxies that  show no strong warps and no sudden changes of position angle. The method works well for galaxies with intermediate inclination angles and smooth and unperturbed disks, but  
 gives less reliable results 
for galaxies with complicated kinematics and geometries, 
 for instance if 
they have recently gone through a merger event that disturbed the disk.

We find clear evidence for radial inflow in NGC~2403 with an amplitude of $15\;$km~s$^{-1}$ corresponding to an H\,{\sc i} mass flow rate of 3~M$_{\odot}\,\mathrm{yr}^{-1}$. Substantial inwards mass flux through the outer disk is also detected in NGC~3198 with values up to 2~M$_{\odot}\,\mathrm{yr}^{-1}$, and to a lower degree in NGC~2903 and NGC~7331 with values around and below 1~M$_{\odot}\,\mathrm{yr}^{-1}$. For NGC~6946 we infer a surprisingly large inflow rate of 15~M$_{\odot}\,\mathrm{yr}^{-1}$, but this measurement is affected by the low inclination of the galaxy and is likely to be overestimated.

We compare our radial mass flux profiles to star formation rate profiles based on GALEX FUV imaging and to SFR measurements from the literature.
We find that 
the rate at with gas gets transported through the disk is sufficient to explain the current star formation activity and if this is indeed representative of the long-term average, than the accretion of new material onto the outer disk of these galaxies and its subsequent inwards transport through the disk is sufficient to maintain a steady state over secular timescales. 

However, a detailed comparison of the radial distribution of star formation and mass inflow usually deviates substantially from the naive assumption of a stationary state with a one-to-one relation between inflow and star formation across galaxy disks. This is indicative of additional mass transport processes acting within the galaxies, for example gas ejection by stellar winds, often referred to as galactic fountains.

For NGC~5055 as well as for NGC~2841 and NGC~925 we find that H\,{\sc i} gas flows outwards in the outer disk. In NGC~5055, there is evidence of a recent encounter with a dwarf galaxy. We think that this is the cause for its perturbed disk geometry, and speculate that the angular momentum delivered to the disk by this encounter drives the inferred outward motion. Indications of a recent merger with a gas-rich  dwarf galaxy are also reported for NGC~925. We therefore consider the encounter with a dwarf galaxy as possible explanation for the peculiar properties of all three objects.

Nearly all the galaxies in our sample show some degree of asymmetry like lobes, streams, or other types of disk perturbations. Complex kinematic structures thus appear quite common among large spiral galaxies and unperturbed disks are likely to be an exception. To properly model the velocity fields of the observed H\,{\sc i} disks, in particular of their outer regions, where these perturbations seem more pronounced, the employed fitting has to be able to adapt to these asymmetries. It would therefore be highly desirable to include as many Fourier components as possible in the simultaneous fit of the H\,{\sc i} velocity field, which is a difficult task:  the components of the Fourier decomposition of the field are highly degenerate and going beyond the second order is not practical. Clearly, more sensitive observations that cover larger areas especially of the out disk will help to improve our ability to study the H\,{\sc i} velocity field, however, the general limitations of the purely kinematic analysis method remains. Doing full justice to the complex velocity patterns observed in these disk galaxies remains a challenging task.

We conclude, that there are indeed radial inflows in the disks of galaxies present which can easily be of sufficient magnitude to sustain star formation in the inner parts of galactic disks. However, the situation appears more complex than simply linking the inferred inflow rates to measured star formation rates: the inflow rates are often up to an order of magnitude larger than the SFR (to some degree a detection bias, since an inflow rate corresponding to the average SFR of $\mathrm{\approx 1\, M_{\odot}\,yr^{-1}}$ in our sample would only be barely detectable with our method) and in every case highly variable throughout the disk. There are a variety of possible explanations like episodic gas accretion, probably related to minor merger events, accretion directly to intermediate radii of the disks or a substantial contribution from internal cycling of gas suggested by the galactic fountain picture.
Without dedicated simulations aiming at a precise modelling of gas kinematics including physical feedback schemes, it will probably be impossible to decide in favour of a particular model.    
While our study based on kinematic data alone could show the existence of radial inflows in disk galaxies, any further investigation will probably not only need superior (deep, extended and good resolution) observational data but also a much broader approach supported by simulations.

\section*{Acknowledgements}

We thank Milan den Heijer, Juergen Kerp and Shahram Faridani for supplying Effelsberg single dish data and stimulating discussions. 
This work made extensive use of the \textit{The HI Nearby Galaxy Survey} (THINGS) database \citep{walter2008}, as well as the NASA/IPAC Extragalactic Database (NED), which is operated by the Jet Propulsion Laboratory, California Institute of Technology, under contract with the National Aeronautics and Space Administration. 
FB acknowledges support from DFG grant BI 1546/1-1.
RSK thanks the Deutsche Forschungsgemeinschaft (DFG) for funding via the SFB 881 {\em The Milky Way System} (subprojects B1, B2, and B8) as well as via the SPP 1573 {\em The Physics of the Interstellar Medium}. 
RSK furthermore acknowledges support from the European Research Council under the European Community's Seventh Framework Programme (FP7/2007- 2013) via the ERC Advanced Grant STARLIGHT (project number 339177). 
WJGdB was supported by the European Commission (grant FP7-PEOPLE-2012-CIG \#333939).

\label{lastpage}


\begin{thebibliography}{99}

\bibitem[\protect\astroncite{Adams et~al.}{2013}]{Adams2013}
Adams, S.~M., Kochanek, C.~S., Beacom, J.~F., Vagins, M.~R.,  Stanek, K.~Z., 2013, ApJ, {778}, 164
  
\bibitem[{Agertz {et~al.}(2009)Agertz, Teyssier, \& Moore}]{Agertz2009}
Agertz O., Teyssier R., Moore B., 2009, MNRAS, 397, L64

\bibitem[\protect\citeauthoryear{Battaglia et al.}{2006}]{Battaglia2006} Battaglia G., Fraternali F., Oosterloo T., Sancisi R., 2006, A\&A, 447, 49

\bibitem[\protect\citeauthoryear{Begeman}{1987}]{Begeman1987}	Begeman K. G., 1987, Ph.D. thesis, Kapteyn Institute, Univ. of Groningen

\bibitem[\protect\citeauthoryear{Bigiel et al.}{2008}]{Bigiel2008}	Bigiel F., Leroy A., Walter F., Brinks E., de Blok W. J. G., Madore B., Thornley M. D., 2008, Astron. J., 136, 2846

\bibitem[\protect\citeauthoryear{Bigiel et al.}{2010a}]{Bigiel2010a}	Bigiel F., Leroy A., Walter F., Blitz L., Brinks E., de Blok W. J. G., Madore, B., 2010, Astron. J., 140, 1194

\bibitem[\protect\citeauthoryear{Bigiel et al.}{2010b}]{Bigiel2010b}	Bigiel F., Leroy A. K., Seibert M., Walter F., Blitz L., Thilker D., Madore B., 2010, ApJ, 720L, 31

\bibitem[\protect\citeauthoryear{Bigiel et al.}{2011}]{Bigiel2011}	Bigiel F., Leroy A., Walter F., Brinks E., de Blok  W. J. G., Kramer  C., Rix  H. W., Schruba  A., Schuster K.-F., Usero A., Wiesemeyer H. W., 2011,	ApJ, 730L, 13

\bibitem[\protect\citeauthoryear{Binney, Dehnen \& Bertelli }{Binney et al.}{2000}]{Binney2000} Binney J., Dehnen W., Bertelli G., 2000, MNRAS, 318, 658

\bibitem[\protect\astroncite{Bird et~al.}{2013}]{Bird2013}	Bird, S., Vogelsberger, M., Sijacki, D., Zaldarriaga, M., Springel, V.,  2013, MNRAS, {429}, 3341

\bibitem[\protect\citeauthoryear{Bosma}{1978}]{Bosma1978}	Bosma A., 1978, Ph.D. thesis, Groningen Univ.

\bibitem[{{Bournaud} {et~al.}(2005){Bournaud}, {Combes}, {Jog}, \&
  {Puerari}}]{bou05} {Bournaud} F., {Combes} F., {Jog} C.~J., {Puerari} I., 2005, A\&A, 438, 507

\bibitem[\protect\citeauthoryear{Brandt}{1960}]{Brandt1960} Brandt J.C., 1960, ApJ, 131, 293

\bibitem[\protect\citeauthoryear{Bregman}{1980}]{Bregman1980}	Bregman J. N., 1980, ApJ, 236, 577

\bibitem[{Ceverino {et~al.}(2010)Ceverino, Dekel, \& Bournaud}]{Ceverino2010} Ceverino D., Dekel A., Bournaud F., 2010, MNRAS, 404, 2151

\bibitem[{Chiappini {et~al.}(2002)Chiappini, Renda, \&  Matteucci}]{Chiappini2002} Chiappini C., Renda A., Matteucci F., 2002, A{\&}A, 395, 789

\bibitem[\protect\citeauthoryear{Chonis et al.}{2011}]{Chonis2011}Chonis T. S., Martinez-Delgado D, Gabany R. J., Majewski S. R., Hill G. J., Gralak R., Trujillo I., 2011, Astron. J., 142, 166

\bibitem[\protect\citeauthoryear{Daddi et al.}{2010}]{Daddi2010} Daddi E., Bournaud F., Walter F., Dannerbauer H., Carilli C. L., Dickinson M., Elbaz D., Morrison G. E., Riechers D., Onodera M., Salmi F., Krips M., Stern D., 2010, Astro. J., 713, 686

\bibitem[\protect\citeauthoryear{de~Blok et al.}{2008}]{deBlok2008} de Blok  J. G., Walter F., Brinks  E., Trachternach  C., Oh  S-H., Kennicutt R. C., 2008,  AJ, 136, 2648

\bibitem[{Dekel {et~al.}(2009)Dekel, Birnboim, Engel, Freundlich, Goerdt,
  Mumcuoglu, Neistein, Pichon, Teyssier, \& Zinger}]{Dekel:2009p1173}
Dekel A., Birnboim Y., Engel G., Freundlich J., Goerdt T., Mumcuoglu M.,
  Neistein E., Pichon C., Teyssier R., Zinger E., 2009, Nature, 457, 451

\bibitem[\protect\astroncite{{Ferri{\'{e}}re}}{2001}]{Ferriere2001}
{Ferri{\'{e}}re}, K.~M., 2001, RMP, {73}, 1031

\bibitem[\protect\citeauthoryear{Fraternali \& Binney}{2006}]{Fraternali2006}	Fraternali F., Binney J. J., 2006, MNRAS, 366, 449

\bibitem[\protect\citeauthoryear{Fraternali \& Binney}{2008}]{Fraternali2008}	Fraternali F., Binney J. J., 2006, MNRAS, 386, 935

\bibitem[{{Fraternali} {et~al.}(2005){Fraternali}, {Oosterloo}, {Sancisi}, \&
  {Swaters}}]{fra05}
{Fraternali} F., {Oosterloo} T.~A., {Sancisi} R., {Swaters} R., 2005, in
  Astronomical Society of the Pacific Conference Series, Vol. 331, Extra-Planar
  Gas, {R.~Braun}, ed., pp.\ 239
  
\bibitem[\protect\citeauthoryear{Fraternali et al.}{2002}]{Fraternali2002}	Fraternali F., van Moorsel G., Sancisi R., Oosterloo T., 2002, AJ, 123, 3124

\bibitem[\protect\citeauthoryear{Genzel et al.}{2010}]{Genzel2010}	Genzel R., Tacconi L. J., Gracia-Carpio J., Sternberg, A., Cooper M. C., Shapiro K., Bolatto A., Bouche N., Bournaud F., Burkert A., Combes F., Comerford J., Cox P., Davis M., Schreiber N. M. F\"orster, Garcia-Burillo S., Lutz D., Naab T., Neri R., Omont A., Shapley A., Weiner B., 2010, MNRAS, 407, 2091

\bibitem[\protect\astroncite{Heald et~al.}{2011}]{Heald2011}Heald G., Józsa G., Serra P., Zschaechner L., Rand R., Fraternali F., Oosterloo T., Walterbos R., Jütte E., Gentile G., 2011, A\&A, 526, 118
	
\bibitem[\protect\astroncite{Hopkins et~al.}{2008}]{Hopkins2008} Hopkins, A.~M., McClure-Griffiths, N.~M.,  Gaensler, B.~M.,2008, ApJ, {682},  L13 

\bibitem[{{Jiang} \& {Binney}(1999)}]{jia99} {Jiang} I., {Binney} J., 1999, MNRAS, 303, L7

\bibitem[\protect\astroncite{Kalberla}{2003}]{Kalberla2003} Kalberla, P. M.~W., 2003, ApJ, {588}, 805
  
\bibitem[\protect\citeauthoryear{Kerp et al.}{2011}]{Kerp2011} Kerp J., Winkel B., Ben Bekhti N., Fl\"oer L., Kalberla P. M. W., 2011, Astron. Nachrichten, 332, 637

\bibitem[\protect\citeauthoryear{Kuzio de Naray et al.}{2012}]{Kuzio2012} Kuzio de Naray R., Arsenault C. A., Spekkens K., Sellwood J. A., McDonald M., D. Simon J. D., Teuben P., 2012, MNRAS, 427, 2523

\bibitem[\protect\astroncite{{Klessen} \& {Hennebelle}}{2010}]{Klessen2010}
{Klessen}, R.~S.,  {Hennebelle}, P., 2010, A\&A, {520}, A17 

\bibitem[{{Kornreich} {et~al.}(2002){Kornreich}, {Lovelace}, \&   {Haynes}}]{kor02} {Kornreich} D.~A., {Lovelace} R.~V.~E., {Haynes} M.~P., 2002, ApJ, 580, 705
  
\bibitem[\protect\astroncite{Leroy et~al.}{2012}]{Leroy2012}
Leroy, A.~K., Bigiel, F., de~Blok, W. J.~G., Boissier, S., Bolatto, A., Brinks,
  E., Madore, B., Munoz-Mateos, J.-C., Murphy, E., Sandstrom, K., Schruba, A.,
   Walter, F., 2012, AJ, {144}, 3

\bibitem[\protect\citeauthoryear{Leroy et al.}{2008}]{Leroy2008}Leroy Adam K., Walter F., Brinks E., Bigiel F., de Blok W. J. G., Madore B., Thornley M. D., 2008, Astron. J., 136, 2782

\bibitem[\protect\citeauthoryear{Leroy et al.}{2009}]{Leroy2009}Leroy A. K., Walter F., Bigiel F., Usero A., Weiss A., Brinks E., de Blok W. J. G., Kennicutt R. C., Schuster K-F., Kramer C., Wiesemeyer H. W., Roussel H., 2009, AJ, 137, 4670

\bibitem[\protect\astroncite{Leroy et~al.}{2013}]{Leroy2013}
Leroy, A.~K., Walter, F., Sandstrom, K., Schruba, A., Munoz-Mateos, J.-C.,
  Bigiel, F., Bolatto, A., Brinks, E., de~Blok, W. J.~G., Meidt, S., Rix,
  H.-W., Rosolowsky, E., Schinnerer, E., Schuster, K.-F.,  Usero, A., 2013, AJ, {146}, 19 
  
\bibitem[{Linsky(2003)}]{Linsky2003} Linsky J.~L., 2003, Space Sci Rev, 106, 49

\bibitem[{Lubowich {et~al.}(2000)Lubowich, Pasachoff, Balonek, Millar,
  Tremonti, Roberts, \& Galloway}]{Lubowich2000} Lubowich D.~A., Pasachoff J.~M., Balonek T.~J., Millar T.~J., Tremonti C.,  Roberts H., Galloway R.~P., 2000, Nature, 405, 1025
  
\bibitem[\protect\citeauthoryear{van der Marel \& Franx}{1993}]{vanderMarel1993}{van der Marel R. P., Franx M., 1993, ApJ, 407, 525}

\bibitem[\protect\astroncite{{Mac~Low} \& {Klessen}}{2004}]{MacLow2004} {Mac Low}, M.-M.,  {Klessen}, R.~S., 2004, RMP, {76}, 125 
  
\bibitem[\protect\citeauthoryear{Marinacci, Pakmor \& Springel}{Marinacci et al.}{2014}]{Marinacci2013}Marinacci F., Pakmor R., Springel V., 2014, MNRAS, 437, 1750

\bibitem[\protect\citeauthoryear{Martinez-Delgado et al.}{2010}]{Martinez-Delgado_2010}Martinez-Delgado D., Gabany R. J., Crawford K., Zibetti S., Majewski S. R., Rix H.-W., Fliri J., Carballo-Bello J. A., Bardalez-Gagliuffi D. C., Penarrubia J., Chonis T. S., Madore B., Trujillo I., Schirmer M., McDavid D. A., 2010, Astron. J., 140, 962

\bibitem[\protect\citeauthoryear{Meidt et al.}{2013}]{Meidt2013} Meidt S. E., Schinnerer E., Garcia-Burillo S., Hughes A., Colombo D., Pety J., Dobbs C. L., Schuster K. F., Kramer C., Leroy A. K., Dumas G., Thompson T. A., 2013, AJ, 779,45

\bibitem[{{Miller} {et~al.}(2009){Miller}, {Bregman}, \& {Wakker}}]{mil09} {Miller} E.~D., {Bregman} J.~N., {Wakker} B.~P., 2009, ApJ, 692, 470

\bibitem[\protect\astroncite{Naab \& Ostriker}{2006}]{Naab2006} Naab, T.,  Ostriker, J.~P., 2006, MNRAS, {366}, 899
  
\bibitem[{{Ostriker} \& {Binney}(1989)}]{ost89} {Ostriker} E.~C., {Binney} J.~J., 1989, MNRAS, 237, 785

\bibitem[{Ostriker \& Tinsley(1975)}]{Ostriker1975} Ostriker J.~P., Tinsley B.~M., 1975, ApJ, 201, L51

\bibitem[{Peek(2009)}]{Peek2009} Peek J. E.~G., 2009, ApJ, 698, 1429

\bibitem[{{Pflamm-Altenburg} \& {Kroupa}(2009)}]{Pflamm2009}
{Pflamm-Altenburg}, J., {Kroupa}, P., 2009, ApJ, 706, 516

\bibitem[\protect\citeauthoryear{Popping, Perez \& Zurita}{Popping et al.}{2010}]{Popping2010}Popping G., Perez I., Zurita A., 2010, A\&A, 521, 13

\bibitem[{Prochaska \& Wolfe(2009)}]{Prochaska2009} Prochaska J.~X., Wolfe A.~M., 2009, ApJ, 696, 1543

\bibitem[\protect\citeauthoryear{Prochaska et al.}{2011}]{Prochaska2011}	Prochaska J. X., Weiner B., Chen H.-W., Mulchaey J., Cooksey K., 2011, ApJ, 740, 91

\bibitem[\protect\citeauthoryear{Rogstad, Lockhart \& Wright}{Rogstad}{1974}]{Rogstad1974}	Rogstad D. H., Lockhart I. A., Wright M. C. H., 1974, ApJ, 193, 309

\bibitem[\protect\citeauthoryear{Saintonge et al.}{2011}]{Saintonge2011} Saintonge, A., Kauffmann, G., Wang, J., Kramer, C., Tacconi, L. J., Buchbender, C., Catinella, B., Graciá-Carpio, J., Cortese, L., Fabello, S., Fu, J., Genzel, R., Giovanelli, R., Guo, Q., Haynes, M. P., Heckman, T.M., Krumholz, M. R., Lemonias, J., Li, C., Moran, S., Rodriguez-Fernandez, N., Schiminovich, D., Schuster, K., Sievers, A., 2011, MNRAS, 415, 61

\bibitem[\protect\citeauthoryear{Saintonge et al.}{2012}]{Saintonge2012} Saintonge, A., Tacconi, L. J., Fabello, S., Wang, J., Catinella, B., Genzel, R., Graciá-Carpio, J., Kramer, C., Moran, S., Heckman, T. M., Schiminovich, D., Schuster, K., Wuyts, S., 2012, ApJ, 758, 73

\bibitem[\protect\citeauthoryear{Salim et al.}{2007}]{Salim2007} Salim S., Rich R. M., Charlot S., Brinchmann J., Johnson B. D., Schiminovich D., Seibert, M., Mallery R., Heckman T. M., Forster K., Friedman P. G., Martin D. C., Morrissey P., Neff S. G., Small T., Wyder T. K., Bianchi L., Donas J., Lee, Y.-W, Madore B. F., Milliard B., Szalay A.S., Welsh B. Y., Yi S. K., 2007, ApJ, 173, 267

\bibitem[\protect\citeauthoryear{Sancisi et al.}{2008}]{Sancisi2008} Sancisi R., Fraternali F., Oosterloo T., van der Hulst T., 2008, A\&A Review, 15, 189

\bibitem[\protect\citeauthoryear{Schoenmakers, Franx \& de Zeeuw}{Schoenmakers et al.}{1997}]{Schoenmakers1997} Schoenmakers R. H. M., Franx M., de Zeeuw P. T., 1997, MNRAS, 292, 349

\bibitem[\protect\citeauthoryear{Schlafly \& Finkbeiner}{2011}]{Schlafly2011} Schlafly E., Finkbeiner, D. P., 2011, ApJ, 737, 103
	
\bibitem[\protect\citeauthoryear{Sellwood \& Sanchez}{2010}]{Sellwood2010} Sellwood J. A., Sanchez R. Z., 2010, MNRAS, 404, 1733

\bibitem[\protect\citeauthoryear{Sellwood \& Spekkens}{2015}]{Sellwood2015} Sellwood J. A., Spekkens, K., 2015, arXiv:1509.07120

\bibitem[\protect\citeauthoryear{Shapiro \& Field}{1976}]{Shapiro1976}Shapiro P. R., Field G. B., 1976, ApJ, 205, 762

\bibitem[\protect\astroncite{Shetty, Clark,  Klessen}{2014}]{Shetty2014b} Shetty R.,  Clark P.~C., Klessen R.~S., 2014, MNRAS, {442}, 2208

\bibitem[Shetty et~al.(2013)]{Shetty2013} Shetty, R., Kelly, B.~C.,  Bigiel, F., 2013, MNRAS, {430}, 288 

\bibitem[Shetty et~al.(2014)]{Shetty2014a} Shetty, R., Kelly, B.~C., Rahman, N., Bigiel, F., Bolatto, A.~D., Clark, P.~C., Klessen, R.~S.,  Konstandin, L.~K., 2014, MNRAS, {437}, L61

\bibitem[\protect\citeauthoryear{Spekkens \& Sellwood}{2007}]{Spekkens2007} Spekkens K., Sellwood J. A., 2007, ApJ, 664, 204

\bibitem[\protect\citeauthoryear{Trachternach et al.}{2008}]{Trachternach2008}	Trachternach C., de Blok W. J. G., Walter F., Brinks, E., Kennicutt R. C. Jr., 2008, Astron. J., 136, 2720-2760

\bibitem[{{van Woerden} \& {Wakker}(2004)}]{van04}
{van Woerden} H., {Wakker} B.~P., 2004, in Astrophysics and Space Science
  Library, Vol. 312, High Velocity Clouds, {H.~van Woerden, B.~P.~Wakker,
  U.~J.~Schwarz, \& K.~S.~de Boer }, ed., pp.\ 195
  
\bibitem[\protect\citeauthoryear{Walter et al.}{2008}]{walter2008} Walter F., Brinks E., de Blok W. J. G., Bigiel F., Kennicutt R. C. Jr., Thornley M. D., Leroy A., 2008, Astron. J., 136, 2562 

\bibitem[\protect\citeauthoryear{Wong, Blitz \& Bosma}{Wong et al.}{2004}]{Wong2004} Wong T., Blitz L., Bosma A., 2004, ApJ, 605, 183

\bibitem[{{Zaritsky} \& {Rix}(1997)}]{Zaritsky1997} {Zaritsky} D., {Rix} H., 1997, ApJ, 477, 118
\end{thebibliography}
\end{document}